\documentclass[aip,jcp,reprint]{revtex4-1}

\usepackage{graphicx}
\usepackage{amsmath}
\usepackage{bm}
\usepackage{mathtools}
\usepackage{color}
\usepackage{acro}
\DeclareAcronym{cpu}{
  short = CPU ,
  long  = Central Processing Unit
}

\DeclareAcronym{gpu}{
  short = GPU ,
  long  = Graphics Processing Unit
}

\DeclareAcronym{plcg}{
    short = PLCG,
    long = preconditioned linear conjugate gradient
}

\DeclareAcronym{maze}{
    short = MaZe,
    long = Mass-Zero constrained dynamics
}

\DeclareAcronym{pmaze}{
    short = Poisson MaZe,
    long = Poisson Mass-Zero constrained dynamics,
    first-style = short
}

\DeclareAcronym{fmm}{
  short = FMM,
  long  = Fast Multipole Method
}

\DeclareAcronym{lmf}{
  short = LMF,
  long  = Local Molecular Field Theory
}

\DeclareAcronym{msm}{
  short = MSM,
  long  = Multilevel Summation Method
}

\DeclareAcronym{mg}{
  short = MG,
  long  = multigrid
}

\DeclareAcronym{lmd}{
  short = LMD,
  long  = Local Molecular Dynamics
}

\DeclareAcronym{mc}{
  short = MC,
  long  = Monte Carlo
}
\DeclareAcronym{fft}{
    short = FFT,
    long = Fast Fourier transform
}

\DeclareAcronym{pme}{
    short = PME,
    long = Particle Mesh Ewald
}

\DeclareAcronym{p3m}{
    short = P3M,
    long = Particle-Particle Particle-Mesh
}

\DeclareAcronym{dof}{
    short = DoF,
    long = degrees of freedom
}

\DeclareAcronym{md}{
    short = MD,
    long = molecular dynamics
}

\DeclareAcronym{pbc}{
short=PBC,
long= periodic boundary conditions
}

\DeclareAcronym{sor}{
short=SOR,
long=Successive Over-Relaxation
}

\DeclareAcronym{vacf}{
short=VACF,
long=velocity autocorrelation function
}

\DeclareAcronym{msd}{
short=MSD,
long=mean squared displacement
}
\acsetup{first-style=long-short}

\definecolor{ao}{rgb}{0.0, 0.5, 0.0}
\begin{document}

\title{Mass-Zero constrained molecular dynamics for electrostatic interactions}

\author{Federica Troni}
\affiliation{Centre Européen de Calcul Atomique et Moléculaire (CECAM), Ecole Polytechnique Fédérale de Lausanne, 1015 Lausanne, Switzerland}
\author{Davide Grassano}
\affiliation{Centre Européen de Calcul Atomique et Moléculaire (CECAM), Ecole Polytechnique Fédérale de Lausanne, 1015 Lausanne, Switzerland}
\author{Jayashree Narayan}
\affiliation{Centre Européen de Calcul Atomique et Moléculaire (CECAM), Ecole Polytechnique Fédérale de Lausanne, 1015 Lausanne, Switzerland}
\affiliation{National Centre for Computational Design and Discovery of Novel Materials (MARVEL), Ecole Polytechnique Fédérale de Lausanne,CH-1015 Lausanne, Switzerland}
\affiliation{Indian Institute for Science Education and Research, Mohali, Knowledge city, Sector 81, SAS Nagar, Punjab 140306, India}
\author{Benoît Roux}
\affiliation{Department of Chemistry, University of Chicago, Chicago, Illinois 60637, United States}
\author{Sara Bonella}
\email[]{sara.bonella@epfl.ch}
\affiliation{Centre Européen de Calcul Atomique et Moléculaire (CECAM), Ecole Polytechnique Fédérale de Lausanne, 1015 Lausanne, Switzerland}
\affiliation{National Centre for Computational Design and Discovery of Novel Materials (MARVEL), Ecole Polytechnique Fédérale de Lausanne,CH-1015 Lausanne, Switzerland}

\begin{abstract}
Optimal exploitation of supercomputing resources for the evaluation of electrostatic forces remains a challenge in molecular dynamics simulations of very large systems. The most efficient methods are currently based on particle-mesh Ewald sums and achieve semi-logarithmic scaling in the number of particles. These methods solve the problem in reciprocal space, requiring extensive use of \acp{fft}. While highly efficient in many contexts, \acp{fft} may encounter scalability challenges at very large processor counts due to their communication requirements. To mitigate these problems, the development and scalable coding of real-space approaches to solve the Poisson equation on a grid is an active field of research. In this work, we introduce a novel real-space approach that provides some advantages over alternatives. Our method exploits an extended Lagrangian in which the values of the field at the grid points are treated as auxiliary variables of zero inertia and the discretized Poisson equation is enforced as a dynamical constraint. The solution of the constraints leads to a linear system - different from those appearing in other real-space approaches - that can be efficiently solved via state-of-the-art iterative methods. The method inherits the numerical scaling of the adopted iterative solver, e.g. linear with a \ac{mg} approach, but converges with fewer cycles. We analyze this approach considering realistic simulations of molten NaCl that validate its ability to reproduce structural and transport properties. Using this non-trivial benchmark, we demonstrate linear scaling and illustrate some features of our algorithm.
\end{abstract}

\maketitle

\section{Introduction}

\Ac{md} simulations based on classical force fields are fundamental for investigating the properties of complex molecular systems.\cite{Allen-Tildesley} In classical \ac{md} simulations, the trajectories of interacting atoms are computed through the discretization of Newton’s laws of motion, with interatomic forces derived from empirical potentials that describe atom-atom interactions. Among these interactions, electrostatic forces pose a computational challenge due to the long-range nature of Coulomb’s law. Historically, long-range electrostatic interactions were often neglected in macromolecular simulations through the use of artificial non-bonded cutoffs. However, studies have demonstrated that such truncations introduce significant artifacts, emphasizing the necessity of accurately incorporating long-range electrostatic interactions. 
Advances in computational power and the development of more efficient algorithms have addressed many of the challenges associated with long-range electrostatics. Perhaps the most widely used approach is the \ac{pme} technique,\cite{Darden:1993, Essmann:1995, Darden-1997} which provides very accurate and computationally efficient results by handling the long-range part of electrostatic interactions in reciprocal space. This method has been adopted by the main simulation codes, including CHARMM,\cite{hwang2024charmm} AMBER,\cite{salomon2013overview} GROMACS,\cite{abraham2015gromacs} NAMD,\cite{NAMD:2020} LAMMPS,\cite{thompson2022lammps} and OpenMM,\cite{eastman2017openmm} and achieves semi-logarithmic scaling in the number of particles. The approach, however, relies heavily on \acf{fft}, whose multidimensional communication patterns can pose challenges for massively parallel systems due to the associated data exchange\cite{Ayala:2022, Zhou:2022}.
To overcome these limitations, several alternative strategies have been developed, each coming with specific pros and cons \cite{George:2022}. Approximate approaches such as the Fast Multipole Method \cite{Gumerov:2004}, Wolf summation \cite{Fennell:2006}, and Local Molecular Field Theory \cite{Rodgers:2008} achieve efficiency by introducing truncations or mean-field corrections, but their accuracy can deteriorate in heterogeneous or interfacial systems. On the other hand, tree-based methods, \cite{Dehnen:2002, Boateng-2019} though capable of achieving linear scaling, face difficulties in handling periodic boundary conditions and depend on empirical parameters that limit their accuracy at high precision. Multilevel summation approaches, \cite{Skeel:2002, Hardy:2014} in contrast, provide systematically improvable accuracy but remain slightly less precise than \ac{pme} and deliver performance benefits primarily in very large or highly parallel simulations.
Among mesh-based strategies, the \ac{p3m}\cite{Hockney:1981} algorithm is a particularly relevant example, that combines explicit short-range interactions with a mesh-based evaluation of the long-range contribution. In the standard formulation of \ac{p3m}, the long-range electrostatics is obtained by solving the Poisson equation in reciprocal space using FFTs. However, real-space variants have also been explored,\cite{Beckers:1998, Sagui:2001} where the reciprocal-space treatment is replaced by iterative or \acf{mg} solvers.  
These approaches have several appealing features: they are mathematically exact and systematically improvable, with accuracy controlled solely by grid resolution and solver tolerance. 
The main limitation of real-space solvers is their iterative nature: although each iteration is inexpensive and involves only local communication, convergence requires multiple steps, making them slower overall than “one-shot” \ac{fft}-based methods. An interesting alternative is provided by a local molecular dynamics formulation, in which the electric field evolves as an auxiliary degree of freedom constrained by Gauss’s law, thereby generating the correct Coulomb interaction with strictly local operations and linear scaling.\cite{Rottler:2004}  \\
In this paper, we take some of these ideas further by considering the problem from an original point of view and introducing a method that presents some numerical advantages compared to previous Poisson solvers. In analogy with these solvers, the approach can be coupled with standard \ac{p3m} algorithms or used directly as a complete solver for electrostatics. In the following, we shall focus on this case. Our approach is based on the \ac{maze} framework. This general and versatile framework was first introduced in the early 1980s to study the rotation-translation coupling in diatomic molecules \cite{Ryckaert:1981}. 
More recently, it has been further developed to treat diverse problems in which physical degrees of freedom interact via parameters subjected to specific conditions. In \ac{maze}, the parameters are promoted to the role of auxiliary degrees of freedom, and the evolution equations of the extended system are determined
via a Lagrangian in which the conditions are
imposed as constraints. The constrained dynamics is solved numerically using standard approaches, such as the SHAKE algorithm. \cite{ryckaert:1977} In the limit of zero mass (inertia) for the auxiliary degrees of freedom, this approach yields an unbiased and efficient dynamics for the physical degrees of freedom that satisfies, by construction, the conditions for the parameters. Applications of \ac{maze} include classical polarizable models~\cite{coretti:2018b, girardier:2021}, constant potential electrochemical simulations~\cite{coretti:2020b, MW:2020}, and \textit{first principles} \ac{md}~\cite{bonella:2020,coretti:2022}. \\
In this work, we present an original extension of the approach showing that it can be used for the calculation of the electrostatic potential and demonstrating its ability to accelerate the convergence of existing Poisson solvers. The general \ac{maze} framework is adapted to this specific problem by identifying the values of the discretized electrostatic potential at the grid points as the auxiliary dynamical variables of zero inertia, and enforcing the discretized Poisson equation as the dynamical constraint. The approach differs from previous adaptations of \ac{maze} in two relevant aspects. Firstly, the form of the extended Lagrangian is engineered specifically for this case. Secondly, in contrast to previous applications that adopted, for example, a non linear conjugate gradient minimization in the SHAKE solution of the constraints, here we employ a \ac{mg} solver that proves much more effective for the specific problem at hand. To underline the specificity of this novel \ac{maze} branch, we shall indicate it in the following as \ac{pmaze}. It should be noted that, even though it exploits MG and ultimately delivers the electrostatic potential, \ac{pmaze} achieves this goal by solving an auxiliary problem that differs from the standard real-space discretization of the Poisson equations, and enables higher computational efficiency.

The paper is organized as follows. After providing the derivation of \ac{pmaze}, we test it simulating a realistic molten salt model. We show that the method can accurately reproduce benchmark structural and transport properties. We also analyze the computational performance of \ac{pmaze} and demonstrate linear scaling of the algorithm with the number of degrees of freedom. Finally, useful theoretical and algorithmic features of the method such as full time reversibility and stability with ionic-driven time step are discussed. The two Appendixes further illustrate some algorithmic choices using a simple toy model (Appendix A) and considering a different iterative solver for the constraints (Appendix B). 
\section{Theory}

Consider $N_p$ point particles described by the set of coordinates $\{\bm{r}_\beta\}$, $\beta=1,...,N_p$, and the charge density $\rho(\bm{r};\{\bm{r_\beta}\})=\sum_{\beta=1}^{N_p}Q_{\beta}\delta(\bm{r}-\bm{r}_{\beta})$, where $Q_\beta$ is the charge of particle $\beta$.
The classical evolution equations for the system are:
\begin{equation}
\label{eq:NewtonPhyDoF}
    m_\alpha \ddot{\bm{r}}_\alpha = -\nabla_{\bm{r}_\alpha}V_c(\{\bm{r}_\beta\}) - \nabla_{\bm{r}_\alpha}{\bar V}(\{\bm{r}_\beta\}) 
\end{equation}
with $\alpha = 1, \cdots, N_p$. In the equation above, ${\bar V}(\{\bm{r}_\beta\})$ represents the non-electrostatic interactions, while 
\begin{equation}
\label{eq:CoulombEnergy}
\begin{split}
    V_c(\{\bm{r}_\beta\}) &= \frac{1}{2}\int d\bm{r} \rho(\bm{r};\{{\bm r}_\beta\}) \phi(\bm{r})
    \\
    &=\frac{1}{2} \sum^{N_P}_{\beta=1}Q_\beta\phi({\bm r}_\beta)
\end{split}
\end{equation}
is the Coulomb potential energy, with the electrostatic potential, $\phi(\bm{r})$, satisfying the Poisson equation:
\begin{equation}
    \bm{\nabla}^2 \phi(\bm{r}) = -4\pi \rho(\bm{r};\{\bm{r}_\beta\})
\end{equation}
Following standard real-space approaches, the Poisson equation is discretized on a 3D grid of $\mathcal{N}$ points. 
The equation is thus reformulated as\cite{Im:1998} :
\begin{equation}
     \bm{M}\bm{\phi} = -4\pi \frac{\bm{q}}{h}
     \label{eq:linearproblem}
\end{equation}
where $\bm{M}/ h^2$ is the matrix representing the discretized Laplacian\footnote{The, standard, form of the matrix $\bm{M}$, is obtained by first considering a discretization on a grid with indices $i, j, k = 0, \ldots, N - 1$, where $N$ denotes the number of grid points along each Cartesian direction. The discrete representation of the Laplacian and the mapping $n = i + jN + kN^2$ are then used to express the matrix with only two indices $n,m = 0,...,\mathcal{N}-1$, where $\mathcal{N} = N^3$. $M_{nm}= -6\delta_{nm} + \delta_{n(m-1)} + \delta_{n(m+1)} + \delta_{n(m-N)} +  \delta_{n(m+N)} + \delta_{n(m-N^2)} +\delta_{n(m+N^2)}$, with appropriate provisions to account for periodic boundary conditions. We recall, for future convenience, that this matrix is symmetrical and has only seven non-zero entries for row.}, and $\bm{\phi}=\{\phi_n\}$ and  and $\bm{q}/h^3=\{q_n/h^3\}$ are the $\mathcal{N}$-dimensional vectors representing the electrostatic potential, and the vector of the fractional charge densities at the grid points, respectively. Equation~\ref{eq:linearproblem} is the linear problem typically tackled iteratively by Poisson solvers. As it is well known, the iterative solution requires an initial guess for the value of the potential and the choice of a solver, e.g. conjugate gradient or - more effectively - \ac{mg}. Common choices for the initial guess include the value of the electrostatic potential at the previous ionic configuration or Taylor series expansions around this value. 
These choices are not dictated by the theory, may vary across different systems, and have a significant impact on the number of iterations required for solver convergence.

In the following, we show that \ac{pmaze} provides a framework to circumvent these issues. 

\begin{figure}[t]
    \centering
    \includegraphics[width=\linewidth]{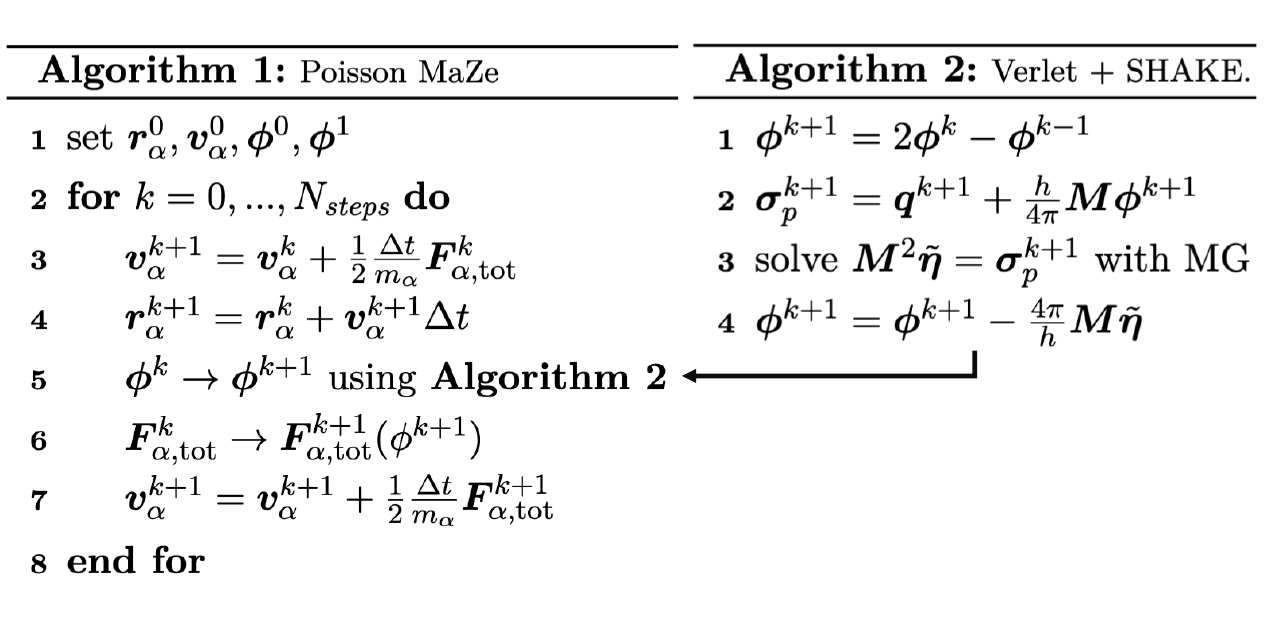}
    \caption{Pseudocode for the \ac{pmaze} algorithm implemented in this work. The index $k$ denotes the time step. The total force $\mathbf{F}_{\alpha, \text{tot}}$ acting on particle $\alpha$ is the sum of the electrostatic force $\bm{F}_\alpha$ (see Equation~\ref{eq:F_nostra}), and a system-dependent non-electrostatic component.}
    \label{fig:algorithm}
\end{figure}

Proceeding in analogy with the standard \ac{maze} approach\cite{Ryckaert:1981,coretti:2018b, bonella:2020, coretti:2020b, girardier:2021, coretti:2022}, we introduce an extended dynamical system in which we add to the physical \ac{dof}, $\{\bm{r}_\beta\}$, a set of auxiliary \ac{dof} associated with the value of the electrostatic potential at each grid point. 
The extended Lagrangian is:
\begin{equation} \label{eq::MaZeLag}
\begin{split}
    \mathcal{L} =& \frac{1}{2}\sum_{\beta=1}^{N_p} m_\beta \dot{\bm{r}}^T_\beta \cdot \dot{\bm{r}}_\beta + \frac{1}{2}\sum_{n = 0}^{\mathcal{N} - 1}\mu \dot{\phi}_n^2 -V(\bm{\phi};\{\bm{r}_\beta\}) +\\& -\bar{V}(\{\bm{r}_\beta\}) -
    \bm{\lambda}^T\cdot \bm{\sigma}
\end{split}
\end{equation}
where we introduced an $\cal{N}$-dimensional vector of constraints $\bm{\sigma}$ (see below) and the corresponding Lagrange multipliers $\bm{\lambda}$.
In the equation above, $\mu$ is the fictitious inertia of the new \ac{dof}. The potential  $V(\bm{\phi};\{\bm{r}_\beta\})$ is defined specifically for the problem at hand and is given by
\begin{equation}
    V(\bm{\phi};\{\bm{r}_\beta\})\coloneq  \frac{h}{8\pi}\bm{\phi}^T\bm{M} \bm{\phi}+  \bm{\phi}^T \cdot \bm{q}  
    \label{eq:potential}
\end{equation} 
where we indicate with the semicolon notation the dependence on the particle's positions via $\bm{\phi}$ and $\bm{q}$.
This form of the potential is chosen for two reasons. Firstly, by substituting the discretized Poisson equation, Equation~\ref{eq:linearproblem}, in first term of Equation \ref{eq:potential}, the potential reduces to a discretized expression of the Coulomb potential energy 
\begin{equation}
\label{eq:energy_discr}
\begin{split}
     \tilde{V}_c(\{\bm{r}_\beta\}) & =\frac{1}{2} \bm{\phi}^T\cdot \bm{q} = \frac{1}{2} \sum_{n=0}^{\mathcal{N} - 1} h^3 \left [\frac{q_n}{h^3}\right ] \phi_n \\ &\xrightarrow[h\rightarrow 0]{} \frac{1}{2}\int d\bm{r} \rho(\bm{r})\phi({\bm{r}})
\end{split}
\end{equation} 
consistent with Equation~\ref{eq:CoulombEnergy}.
Secondly, with this choice the Poisson equation can be cast as a minimization constraint on the potential, highlighting the analogy with previous MaZe applications. In fact, imposing the constraint:
\begin{equation}
\label{eq:Constraints}
\begin{split}
  \bm{\sigma}(\bm{\phi};\{\bm{r}_\beta\}) &= \nabla_{\bm{\phi}^T} V(\bm{\phi};\{\bm{r}_\beta\}) \\
  &= \frac{h}{4\pi} \bm{M}\bm{\phi} +  \bm{q} = 0
  \end{split}
\end{equation}
ensures, after trivial reorganization of the equation, that the auxiliary \ac{dof} satisfy Equation~\ref{eq:linearproblem}.
At this stage, we proceed again in complete analogy with standard \ac{maze}. The evolution equations for the physical \ac{dof} are 
\begin{equation}
\begin{split}
    m_\alpha \ddot{\bm{r}}_\alpha =& -  \nabla_{{\bm r}_\alpha}V(\bm{\phi};\{\bm{r}_\beta\}) - \nabla_{\bm{r}_\alpha}{\bar V}(\{\bm{r}_\beta\}) \\& -\sum_{n=0}^{\mathcal{N} - 1} \lambda_n \nabla_{{\bm{r}}_\alpha}\sigma_n(\bm{\phi};\{\bm{r}_\beta\})
\end{split}
\end{equation}
while, for the auxiliary \ac{dof} we have

\begin{equation}
    \begin{split}
         \ddot{\phi}_n &= -\frac{1}{\mu} \frac {\partial V(\bm{\phi};\{\bm{r}_\beta\})}{\partial \phi_n} -\sum_{m=0}^{\mathcal{N} - 1} \frac{\lambda_m}{\mu} \frac{\partial \sigma_m}{\partial \phi_n} = \\ &= -\frac{1}{\mu} \frac{h}{4\pi}\bm{M}\bm{\lambda}
    \end{split}
    \label{eq:Intermediate}
\end{equation}
In the equation above, the derivative of the potential in the first line is zero due to the constraint conditions. Using the explicit form of the derivative of the constraint and the symmetry of $\bm{M}$ leads to the expression in the second line. At this stage we consider the limit $\mu \rightarrow 0$ of the evolution equations. To preserve finite accelerations of the auxiliary variables, we impose that the Lagrange multipliers $\bm{\lambda}$ go to zero with the same velocity as $\mu$. For notational convenience, we set $\bm{\tilde{\lambda}}=\frac{h}{4\pi}\bm{\lambda}$, and define the (non-null) vector $\bm{\eta}=\lim_{\mu \rightarrow 0} \bm{\tilde{\lambda}}/\mu$. In this zero inertia limit, the evolution equations for the overall system are:
\begin{equation}
\label{eq:MaZeEvolution}
\begin{split}
    m_\alpha \ddot{\bm{r}}_\alpha &= -  \nabla_{{\bm r}_\alpha}V(\bm{\phi};\{\bm{r}_\beta\}) - \nabla_{\bm{r}_\alpha}{\bar V}(\{\bm{r}_\beta\})\\
    \ddot{\bm{\phi} }& = - {\bm{M}} {\bm{\eta}}
\end{split}
\end{equation}
Note that, due to the $\bm{\lambda}$ going to zero condition, the evolution equations for the physical \ac{dof} are unaffected by the constraints. Furthermore, when the grid spacing goes to zero, the discretized Coulomb potential energy becomes the exact one and the equations for the physical \ac{dof} tend to Equation~\ref{eq:NewtonPhyDoF}. 
The dynamical system above therefore provides a consistent approach to determining simultaneously the value of the electrostatic potential and the evolution of the particles. 
\\ \noindent
The extended dynamical system can be solved numerically using standard techniques. In the following, we employ the velocity Verlet algorithm for the ions. To evolve the auxiliary degrees of freedom, subjected only to the constraint force, it is necessary to compute the unknown set of Lagrange multipliers ${\bm{\eta}}$. This is also accomplished straightforwardly using the SHAKE method\cite{ryckaert:1977}. We now summarize this algorithm focusing on some specific aspects related to our problem.This will also make it possible to highlight the differences between \ac{pmaze} and standard Poisson solvers.  
We start by propagating the auxiliary \ac{dof} using the Verlet algorithm rather than velocity Verlet as for the ions. This is a convenient choice because it avoids calculating the (constrained) time-derivatives of the auxiliary variables that are not needed for the propagation of the ions. The Verlet time-step is
\begin{equation}
     \bm{\phi}^{k + 1} = 2\bm{\phi}^k- \bm{\phi}^{k - 1}- \frac{1}{2}\Delta t^2 \bm{M}\bm{\eta}  
\end{equation}
in the equation above, the superscript indicates the time-stepping iteration and $\Delta t$ is the time-step. Following standard SHAKE implementations, we now identify $\bm{\phi}_p^{k + 1} = 2\bm{\phi}^k- \bm{\phi}^{k - 1} $ as the provisional value of the discretized potential, so that $\bm{\phi}^{k + 1}=\bm{\phi}_p^{k + 1}- \frac{1}{2} \Delta t^2\bm{M}\bm{\eta}$. This notation highlights that the time-update depends on the unknown set of Lagrange multipliers, $\bm{\eta}$. These are computed by imposing $\bm{\sigma}(\bm{\phi}^{k + 1}; \{\bm{r}_\alpha\}) = 0$. 
Given the specific form of our constraint (Equation~\ref{eq:Constraints}), this amounts to solving:
\begin{equation}
\label{eq:SHAKE}
\begin{split}
    \frac{h}{4\pi}\bm{M}(\bm{\phi}^{k + 1}_p - \frac{1}{2} \Delta t^2\bm{M}\bm{\eta}) + \bm{q}^{k + 1}&=0\\
\end{split}
\end{equation}
where $\bm{q}^{k+1}$ is the charge density corresponding to the ionic configuration at step $k+1$.  Defining $\bm{\sigma}_p(\bm{\phi}_p^{k + 1}; \{\bm{r}_\alpha\}) = \frac{h}{4\pi}\bm{M}\bm{\phi}^{k + 1}_p+\bm{q}^{k + 1}$ and $\tilde{\bm{\eta}}=\frac{h}{4\pi}\frac{1}{2}\Delta t^2 \bm{\eta}$, the equation above can be expressed as:
\begin{equation}
    \bm{M}^2\tilde{\bm{\eta}} = \bm{\sigma}^{k + 1}_p
\end{equation}
Recall that the matrix $\bm{M}$ is the discretized Laplacian. Note that, since the Verlet update necessitates only the vector  $\bm{M}\tilde{\bm{\eta}}\stackrel{\mathrm{def}}{=}\bm{y}$, we can solve directly
\begin{equation}
\label{eq:SHAKEy}
    \bm{M} \bm{y} = \bm{\sigma}^{k + 1}_p 
\end{equation}
thus avoiding the calculation of the square of matrix. The time-updated values of the discretized electrostatic potential are finally obtained as
\label{eq:SHAKEPhiFinal}
\begin{equation}
    \bm{\phi}^{k + 1} = \bm{\phi}^{k + 1}_p - \frac{4\pi}{h}\bm{y}^*
\end{equation}
where $\bm{y}^*$ is the solution of Equation~\ref{eq:SHAKEy}. This solution is obtained iteratively using a solver of choice, in this work the \ac{mg} method. Figure~\ref{fig:algorithm} shows the pseudocode for the overall algorithm. 
\\
Note that, even though we derived the evolution equations in the (N,V,E) ensemble, the generalization to other ensembles and in particular to (N,V,T) systems, is straightforward. The thermostat (e.g., Langevin) affects only the physical \ac{dof}, leaving the computation of the electrostatic potential in Algorithm~2 unchanged.

To conclude this section, let us point out some relevant differences between \ac{pmaze} and traditional Poisson solvers. Of course, in both cases, the goal is to obtain the electrostatic potential for the assigned ionic configuration. Poisson solvers achieve this by direct numerical solution of Equation~\ref{eq:linearproblem}, a linear system of equations in which the unknown is the discretized potential and the right-hand side is the discretized charge density. In \ac{md} calculations of the ionic trajectory, this equation is solved as separate from the dynamics\footnote{in the context of \textit{first principles} \ac{md}, this would be analogous to the Born-Oppenheimer approach - albeit with a different condition for the non ionic \ac{dof}.}. \ac{pmaze} obtains the potential by propagating it together with the ions via a suitable dynamics. At each step, a putative value for the potential is obtained via Verlet and "corrected" via the constraint force to enforce that the potential satisfies the Poisson equation. The correction requires to solve a linear system of equations, see Equation~\ref{eq:SHAKEy}, where the discretized Laplacian matrix appears as in Equation~\ref{eq:linearproblem} but with different unknowns (the vector $\bf{y}$) and right-hand side ($\bm{\sigma}_p$, qualitatively providing a measure of the difference between the putative and correct values of the potential). Both Poisson solvers and \ac{pmaze} then rely on iterative approaches to solve a (different) linear system of equations. The numerical efficiency of either method hinges on the choice of the iterative scheme. The \ac{mg} approach has emerged as the most effective option, with proven linear scaling for traditional Poisson solvers~\cite{gholami:2016, ibeid:2020}. The same approach can be adopted to solve Equation~\ref{eq:SHAKEy} thus ensuring linear scaling also for \ac{pmaze}, see also Section~\ref{subsec:performance}. The speed of convergence of \ac{mg} depends on the value of the unknown used to initialize the iterative cycle of the solver. As mentioned earlier, in traditional Poisson solvers the value of the electrostatic potential at the previous ionic configuration is often used for this purpose. In \ac{pmaze}, following standard practice in implementations of SHAKE, the iterative cycle is initialized via the value of $\bf{y}^*$ at the previous step. In Section~\ref{sec:Results} we show that the \ac{mg} solution of Equation~\ref{eq:SHAKEy} requires less iterations than the \ac{mg} solution of Equation~\ref{eq:linearproblem}, leading to an overall reduction of computational time.

\subsection{Comments on the calculation of force and energy}
\label{subsec:energy_force_theory}

Before moving to the results, we mention a relevant technical point. The calculation of the force after discretization of the electrostatic potential is an important element of the algorithm. In Equation~\ref{eq:MaZeEvolution}, the gradient of $V(\bm{\phi};\{\bm{r}_\beta\})$ is taken with respect to the particle’s position, $\bm{r}_\alpha$, which typically lies off the grid. Several approaches exist in the literature for taking this derivative of the discretized potential~\cite{Hockney:1981, Gilson:1985, Gilson:1993, Im:1998}. 
Following previous work~\cite{Hockney:1981}, in this paper it is evaluated on the grid using a central difference scheme. For example, the derivative along the $x$ direction becomes:
\begin{equation}
    \frac{\partial \phi_m}{\partial x_\alpha} \quad \longrightarrow \quad D_x\phi_n = \frac{\phi_{n+1} - \phi_{n-1}}{2h} ,
\end{equation}
where $D_x$ indicates the discrete derivative along the $x$ axis, the indices $n+1$ and $n-1$ refer to the neighboring grid points in the $x$ direction relative to grid point $n$. The full $x$ component of the force is then computed as:
\begin{equation}
    F_\alpha^x = -\sum_{n \in NN} q_n D_x\phi_n 
    \label{eq:F_nostra}
\end{equation}
where the sum runs over the $8$ grid points that are nearest neighbours ($NN$) of the position of particle $\alpha$, and $q_n$ is the charge assigned to each point. 
This charge is obtained as $q_n=Q_{\alpha}W(\bm{r}_{\alpha} - \bm{r}_n)$ where $\bm{r}_n$ is the vector that identifies the $n$-th nearest neighbor to particle $\alpha$ on the grid, and $W(\bm{r}_{\alpha} - \bm{r}_n)$ is the cubic B-spline weight function. 
Employing the same weight function $W$ for back-interpolation in the force calculation, and using a central difference estimator for the derivatives of the potential, guarantees that the sum of all forces is zero and, therefore, the total momentum of our system is conserved~\cite{Hockney:1981}. Momentum conservation has the additional advantage of ensuring that terms in the force originating from self-energy contributions are null\cite{Hockney:1981}. 

Self-energy contributions are spurious non-zero terms in the electrostatic energy arising from discretization artifacts, e.g. due to the charge assignment scheme~\cite{Hockney:1981, Sagui:2001,Maggs:2004}. They result in discrepancies between the analytical electrostatic energy $V_c$ (Equation \ref{eq:CoulombEnergy}) and the discretized form $\tilde{V}_c$ (Equation \ref{eq:energy_discr}) that can lead to strong oscillations in the energy and - if no corrective measures are taken - in the numerical forces.

The self-energy oscillations in the energy prevent use of $\tilde{V}_c$ as a practical tool to monitor the correct time evolution of the system. We then propose an alternative potential energy defined from the work performed by the particles:
\begin{equation}
    V^\text{work}_c(t) = - \sum_{\alpha=1}^{N_p}\int^{\bm{r}_\alpha(t)}_{\bm{r}^0_\alpha} \bm{F}_\alpha \cdot d\bm{r}_\alpha 
    \label{eq:E_nostra}
\end{equation}
where $\bf{F}_{\alpha}$ is defined as in Equation~\ref{eq:F_nostra}. Since the definition above builds on a self-energy free force, this estimator is more stable for all grid sizes and can be used in simulations to validate, for example, the choice of the time step. 

To further illustrate the points above we mention also an alternative force definition
\begin{equation}
    \tilde{\bm{F}}_\alpha = -\sum_{n \in NN} Q_\alpha \nabla_{\bm{r}_\alpha} W(\bm{r}_\alpha - \bm{r}_n)\phi_n
    \label{eq:F_Benoit}
\end{equation}
that has been proposed in the literature~\cite{Im:1998}. This force does not conserve momentum and is affected by the oscillations of the self-energy thus highlighting the importance of the choice of an appropriate derivative of the discretized electrostatic potential. In Appendix~\ref{sec:Appendix1} we illustrate in some detail the properties of the different force and energy estimators discussed above using a simple one-dimensional example. 

\section{Results}
\label{sec:Results}
\label{subsec:molten_salt}
In this section, we report results for the MD simulation of a realistic model of molten NaCl to show that our approach provides the correct structural and transport properties for a non-trivial test system. The numerical performance of \ac{pmaze} is also compared with data obtained from direct MG solution of the Poisson equation. We further show that the algorithm conserves (within good numerical accuracy) both energy and total momentum, and satisfies time reversibility and stationarity of time-averages. \\ \noindent
The simulated system consists of $250$ ions ($125 \text{ Cl}^-$ and $125 \text{ Na}^+$), placed in a cubic box of side $L = 20.64\,\text{\AA}$, with a density of $\rho = 1.3793 \text{ g} / \text{cm}^3$. Periodic boundary conditions (PBC) are imposed in all directions.
This set up reproduces the one adopted in previous works~\cite{Galamba:2007, Mouhat:2013} that we use as a benchmark, by comparing the radial pair distribution function for the different species, $g_{IJ}(r)$, ($I,J =$ Na, Cl), and the diffusion coefficients.
Interactions in our reference simulations are described by a Born-Huggins-Meyer potential with Tosi-Fumi parameters \cite{Tosi:1964, Mouhat:2013} .

In this work, we removed the electrostatic contribution from the potential and computed it using \ac{pmaze}. For the rest, we consider a modified potential, where we keep the Born-Huggins overlap repulsion potential, the dipole-dipole and dipole-quadrupole dispersion energies as described in the original Tosi-Fumi:
\begin{equation}
    \bar{V}_{\text{TF}}(r_{\alpha\beta}) =  A_{\alpha\beta}e^{B(s_{\alpha\beta}-r_{\alpha\beta})} - \frac{C_{\alpha\beta}}{r^6_{\alpha\beta}} - \frac{D_{\alpha\beta}}{r^8_{\alpha\beta}}
\end{equation}

In the expression above, $r_{\alpha\beta}$ is the distance between particles $\alpha$ and $\beta$, while $A_{\alpha\beta}, B, s_{\alpha\beta},C_{\alpha\beta}, \mathrm{and} \,\, D_{\alpha\beta}$ are constants. Values for these constants are reported in~\cite{Tosi:1964, Mouhat:2013}.
An equilibration run in the canonical ensemble at target temperature $T=1550$ K preceded the production run that was performed in the microcanonical ensemble. The initial configuration for the equilibration run was obtained by placing the ions on a bcc lattice, with initial velocities sampled from a Maxwell-Boltzmann distribution at the target temperature. Constant temperature runs were performed via Langevin dynamics, solving the equations of motion using the OVRVO integrator~\cite{Sivak:2013}. The Langevin friction parameter, $\gamma$, was tuned to obtain fast thermalization setting $\gamma = 10^{-3}$. For the production, constant energy, run, $\gamma$ was set to zero, reducing OVRVO to the Velocity Verlet algorithm. The equilibration run was carried out for $2.5$ ps, with a time-step of $\Delta t = 0.25$ fs. The same time-step was used for the production run of $\mathcal{T} = 25$ ps. The computation of electrostatic forces was performed on a grid with a spacing of $h = 0.172\,\text{\AA}$ (corresponding to $\mathcal{N} = 120^3$ points). 

\begin{figure}[htbp]
    \centering
    \includegraphics[width=0.95\linewidth]{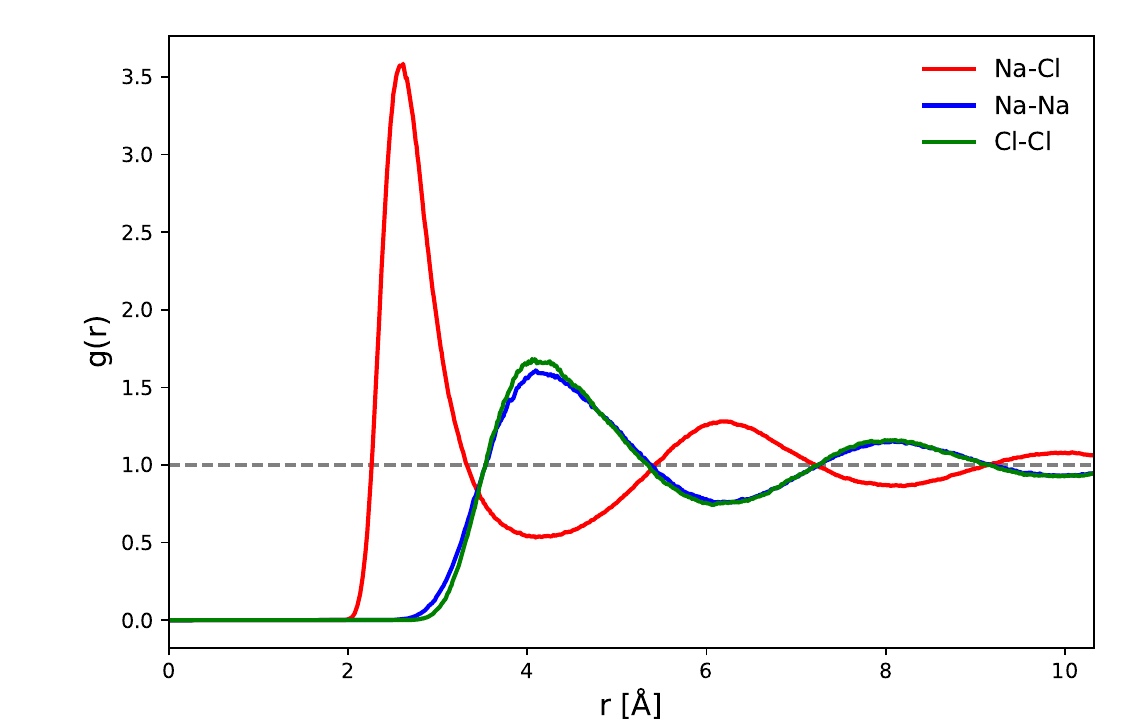}
    \caption{Radial pair distribution function obtained with a simulation of $25$ ps and a grid size $h = 0.172 \ \text{\AA}$.}
    \label{fig:g_r}
\end{figure}

Results for the radial pair distribution functions $g_{IJ}(r)$ are shown in Figure \ref{fig:g_r}, and are in very good agreement with previous simulations\cite{Galamba:2007, Mouhat:2013} . More details are provided in Table \ref{tab:comparison_gr}, where we compare the position and height of the first peak, and the position of the first minimum and of the cutoff (i.e. the distance where the pair distribution function becomes non-null) with those of our reference.

\begin{table}[htbp]
    \centering
    \scriptsize
    \begin{ruledtabular}
    \setlength{\tabcolsep}{4pt} 
    \begin{tabular}{cccccc}
         Method & $g_{IJ}(r)$ & $r_\text{cut}$ [$\text{\AA}$] & $r_\text{max}$ [$\text{\AA}$] & $g_\text{max}$ [$\text{\AA}$] & $r_\text{min}$ [$\text{\AA}$] \\
        \hline
        Poisson MaZe & NaCl  & 1.9 & 2.6 & 3.5 & 4.2 \\
                     & NaNa  & 2.3 & 4.2 & 1.6 & 6.1 \\
                     & ClCl  & 2.6 & 4.1 & 1.7 & 6.1 \\
        \\[-0.9em]
        Galamba et al.\cite{Galamba:2007} & NaCl  & 1.9 & 2.6 & 3.5 & 4.2 \\
                                          & NaNa  & 2.4 & 4.2 & 1.6 & 6.2 \\
                                          & ClCl  & 2.6 & 4.1 & 1.7 & 6.3 \\
        \end{tabular}
    \caption{Analysis of the $g_{IJ}(r)$ parameters, including the cutoff of the curve $r_\text{cut}$, the position $r_\text{max}$ and height of the first maximum $g_\text{max} = g(r_\text{max})$, and the position of the first minimum $r_\text{min}$.}
    \label{tab:comparison_gr}
    \end{ruledtabular}
\end{table}

\begin{figure}[htbp]
    \centering
  \includegraphics[width=0.95\linewidth]{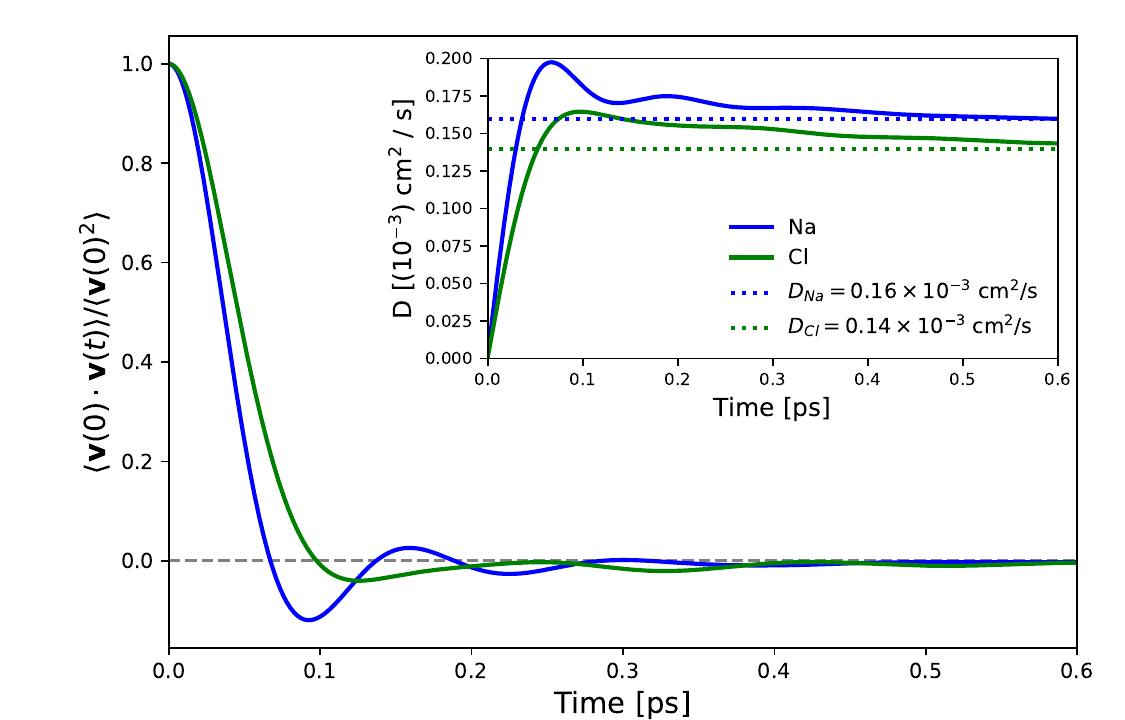}
  \caption{Normalized \acp{vacf} for the two species (main panel) and their integral  (inset) as a function of the integration time, for Na$^+$ (in blue) and Cl$^-$ (in green) at $T=1550 $ K.}
  \label{fig:vacf}
\end{figure}

\noindent
In Figure \ref{fig:vacf}, we show the \acp{vacf} of Na\(^+\) (blue curve) and Cl\(^-\) (green curve), together with their integrals over time. These plots are also in excellent agreement with previous results. The diffusion coefficients for the system were obtained by integrating the \acp{vacf}, as well as from the \ac{msd}. The results are fully consistent with the reference values\cite{Galamba:2007}: 
\[
D_\text{Na} = 0.16 \times 10^{-3} \ \text{cm}^2/\text{s} \quad \text{and} \quad D_\text{Cl} = 0.14 \times 10^{-3} \ \text{cm}^2/\text{s},
\]
and both estimators yield the same values.

\subsection{Performance and scaling}
\label{subsec:performance}

We now discuss the performance of \ac{pmaze} with \ac{mg} solver for the evaluation of the constrains, Equation~\ref{eq:SHAKEy}, and compare it with a direct \ac{mg} solution of the Poisson equation, Equation~\ref{eq:linearproblem}. As mentioned before, direct solutions of the Poisson equation via \ac{mg} scale linearly and have been parallelized very efficiently in state-of-the-art codes\cite{Sagui:2001}. \ac{pmaze} inherits the properties of the solver adopted for the constraints, so, as shown in the following, it will also display linear scaling and profit from parallelization. In addition, the two-step procedure to complete the update of the electrostatic potential (i.e. the Verlet proposal of a putative value followed by the correction enforced via the constraint force) leads to faster convergence of the iterative part of the algorithm and therefore to shorter total runtime.

To set the stage for the results reported in this section, we recall that the \ac{mg} approach is an iterative method that reduces increasingly lower frequency errors by applying a smoothing function on successively coarser grids, and by defining a restriction and prolongation operation that allows to move between a finer and a coarser grid. A sweep of these operations is indicated as a v-cycle that we identify as one iteration. 
A v-cycle starts from the finest grid and consists of a sequence of smoothing steps, restriction of the residual to coarser grids, solution on the coarsest grid, and prolongation of the correction back to finer levels, followed by post-smoothing.
The residual norm is computed at the end of every iteration and the cycle stopped when the residual is below the prescribed tolerance.

The specific \ac{mg} scheme adopted in this work emplys the same operators as that of Sagui and Darden \cite{Sagui:2001}, with Red–Black Gauss–Seidel smoothing, 27-point full-weighting restriction, and trilinear prolongation. We use conjugate gradient as the exact solver for the coarsest grid of the v-cycle. All calculations were performed with 8 processors using an in-house code parallelized with OpenMP\footnote{This implementation is quite certainly not as efficient as \ac{mg} functions in high end community codes, but since here we are interested in relative performance of \ac{pmaze} and \ac{mg} at equal implementation level, it is sufficient for our purposes. We also note that the \ac{mg} function called is the same, with different input, for the direct solution of the Poisson equation and for \ac{pmaze}. This implies that transfer of our approach to community codes requires only the set up of the extended dynamical system, and can directly benefit from existing optimizations.}. 

\subsubsection{Convergence of the iterative cycles for \ac{pmaze} and direct Poisson solver}
\label{subsec:convergence}
We compare the convergence of the iterative solutions for Equations~\ref{eq:linearproblem} and Equation~\ref{eq:SHAKEy} for a set of ionic configurations selected randomly along a long MD trajectory of the NaCl system described in the previous section. The same ionic configuration (or configurations, see below) is used to construct the input for the same MG solver for either system of equations and the path to convergence to the solution is traced via the residual as a function of the iteration step $\nu$. The, standard, definition of the residual for the direct Poisson solution is 
\begin{equation}
   \text{res}_\nu = \max_n \left|\left\{\left(\bm{M}\bm{\phi}^{\nu} + \frac{4\pi}{h}\,\bm{q}\right)_n\right\}\right|.
   \label{eq:res_poisson}
\end{equation}
In this equation, $n$ indexes the components of the vectors in the linear system, $\bm{q}$ is the charge density associated to the assigned ionic configuration, and ${\bm\phi}^{\nu}$ is the value of the associated electrostatic potential  at iteration $\nu$. 
For the \ac{pmaze} system the, standard, definition of the residual is 
\begin{equation}
   \text{res}_\nu = \max_n |\{(\bm{M}\bm{y}^\nu - \bm{\sigma}_p)_n\}|,
    \label{eq:res_poisson_maze}
\end{equation}
where ${\bm y}^{\nu}$ is the value of the Lagrange multipliers (rescaled via the definitions between Equation~\ref{eq:SHAKE} and Equation~\ref{eq:SHAKEy}) at iteration $\nu$, and $\bm{\sigma}_p = \frac{h}{4\pi}\bm{M}\bm{\phi}^{k + 1}_p+\bm{q}^{k + 1}$ is the value of the constraint at computed at the putative potential $\bm{\phi}^{k+1}_p = 2\bm{\phi}^k - \bm{\phi}^{k-1}$ and for the charge density associated to the assigned ionic configuration (the time indexing of these quantities via $k$, omitted in the equations above to simplify the notation, is consistent with the notation in the Theory section). 

As mentioned earlier, the starting (zero iteration) value of the unknown affects the speed to convergence of an iterative solver. For \ac{pmaze}, we use the standard (for SHAKE) assignment to start the iterations: $\bm{y}^0$ is equal to the exit value at the previous time step. For \ac{mg}, we consider two choices. The first, that we indicate as 
\textit{static}, uses as starting value the converged electrostatic potential at the previous time step. The second, that we indicate as \textit{time-propagated}, uses the putative value of the field as computed in the \ac{pmaze}, $\bm{\phi}^{k+1}_p = 2\bm{\phi}^k - \bm{\phi}^{k-1}$. This choice was made to minimize the differences in the initializations between the direct Poisson solver and \ac{pmaze}.

\begin{figure}[htbp]
   \centering
\includegraphics[width=0.95\linewidth]{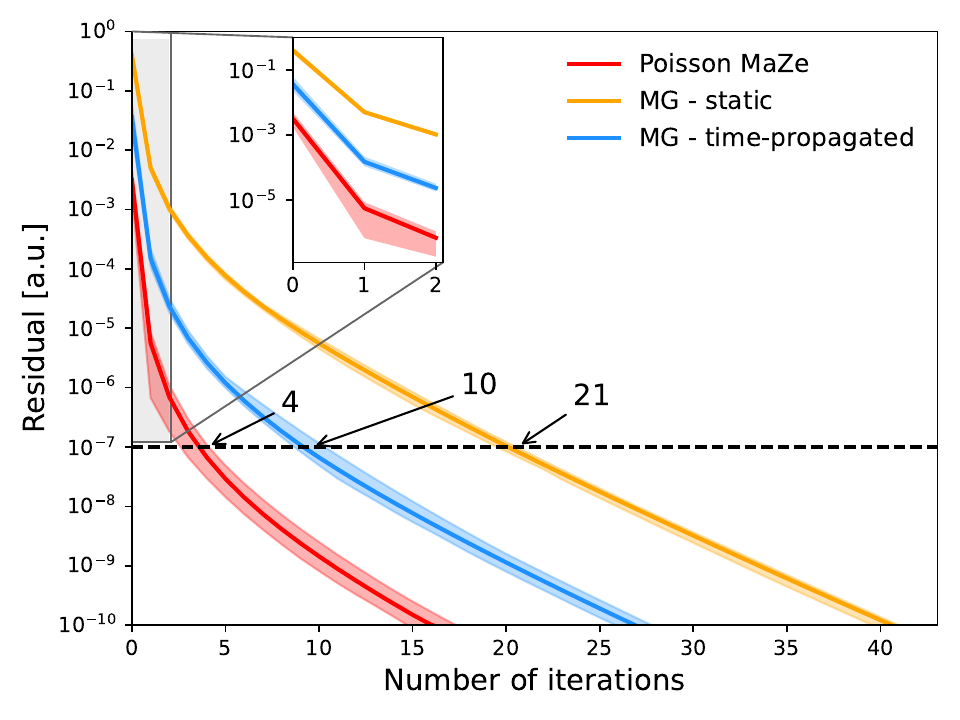}
\caption{Convergence of \ac{pmaze} (red), plain \ac{mg} with \textit{static} (yellow), 
and \textit{time-propagated} (blue) starting value for the iterations. Solid lines show the mean residual across trajectories, and shaded bands indicate the pointwise min–max envelope. Arrows mark the iteration at which the residual crosses $10^{-7}$.}
   \label{fig:tol_vs_v_cycles}
\end{figure}

In Figure~\ref{fig:tol_vs_v_cycles}, we show results for the residuals as a function of the number of iterations, obtained averaging the path to convergence over $10^3$ ionic configurations. In the figure, the solid line indicates the average of the residuals, while the shaded bands represent the variability of the measurements at each iteration over all configurations. The figure shows that the \ac{pmaze} solution (red curve) is faster than the \ac{mg} solutions obtained either with the \textit{static} (yellow curve) or \textit{time-propagated} (blue curve) initialization. For a tolerance of $10^{-7}$ (same as that used in the NaCl calculations), \ac{pmaze} converges in only four iterations (see arrows in Figure~\ref{fig:tol_vs_v_cycles}), whereas the \textit{time-propagated} \ac{mg} scheme requires roughly twice as many iterations, and the \textit{static} \ac{mg} roughly seven times as many. The better performance of \ac{pmaze} is preserved at lower convergence thresholds. The inset of Figure~\ref{fig:tol_vs_v_cycles} highlights the effect of the different choices for the starting value of the iterations, indicating that the combined use of the putative potential and of the previous value of the constraints leads to a measurable numerical advantage for \ac{pmaze}. Crucially, the improvement in the number of iterations to convergence is achieved without incurring a higher per-iteration cost: the average CPU times per iteration are $(4.3 \pm 0.3)\times 10^{-2}$ s for \ac{pmaze}, $(4.8 \pm 0.3)\times 10^{-2}$ s for \ac{mg} with \textit{static} initial guess, and $(4.5 \pm 0.3)\times 10^{-2}$ s for \ac{mg} with the \textit{time-propagated} one. This shows that the cost of a single iteration is dominated by the \acl{mg} structure itself and does not depend significantly on the initial guess strategy or the use of the \ac{maze} superstructure. It also means that the reduced number of iterations required for convergence of \ac{pmaze} results into a shorter total runtime. We also note that the precision of the calculated electrostatic potential is the same, with differences between the values of the fields computed at the end of the iterative cycle typically well below the set convergence threshold. The absolute value of this difference for \ac{pmaze} vs \textit{static} direct solver is of the order of $10^{-10}$, with the value in the case of \ac{pmaze} vs \textit{time-propagated} direct solver is of the order of $10^{-14}$, independent of the set threshold.

\subsubsection{Scaling with system size at constant density}
\begin{figure}[htbp]
   \centering
   \includegraphics[width=0.95\linewidth]{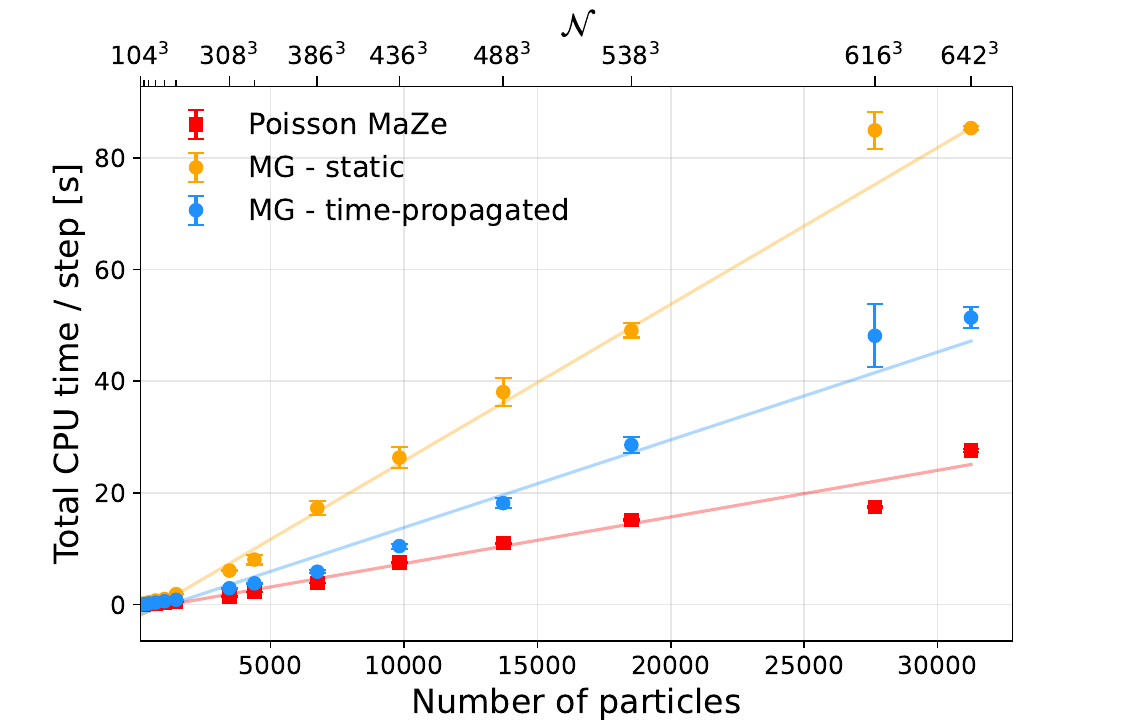}
    \caption{Scaling of the total CPU time per step as a function of the number of particles for \ac{pmaze} and \ac{mg} with two different initial guesses (\textit{static} in yellow, \textit{time-propagated} in blue). The corresponding number of grid points is reported on the top axis. All methods display linear scaling, consistent with the underlying common \ac{mg} solver.}
   \label{fig:scaling_density}
\end{figure}

We now verify numerically the expected linear scaling  with system size of the \ac{mg} based \ac{pmaze}. For comparison, we also report results for the direct \ac{mg} solutions of the Poisson equation with the \textit{static} and \textit{time-propagated} starting schemes introduced in the previous section. All simulations were performed at constant density $\rho = 1.3793 \text{ g} / \text{cm}^3$ and fixed grid spacing $h=0.172\,\text{\AA}$. The number of particles was varied between $N_p=128$  and $N_p=31250$. After an equilibration run, 100 steps were performed with each solver, with an imposed stopping tolerance of $\rm{tol=10^{-7}}$ on the residual, and the CPU time and number of iterations per step averaged over this short trajectory. Figure~\ref{fig:scaling_density}, reports scalings with the number of particles. The top axis of the figure indicates the number of grid points associated to each $N_p$. All runs exhibit linear scaling as dictated by the \ac{mg}. With this approach, in fact, the cost of a single iteration grows linearly with the number of grid points, while the number of iterations remains essentially constant, leading overall to $\mathcal{O}(N_p)$ scaling since the number of grid points increases proportionally to $N_p$ \cite{briggs:2000, Sagui:2001}.
Consistent with the results of the previous subsection, the \ac{pmaze} runs (red data points in Figure~\ref{fig:scaling_density}) show a reduction in the overall CPU time for all system sizes. We quantify this speedup by fitting the total CPU time per step against the system size using a weighted linear regression. The slopes of these fits (shown as solid lines in the figure) are $m_\text{PM} = (8.36 \pm 0.53)\times 10^{-4}$ for \ac{pmaze}, $m_\text{MG-s} = (2.81 \pm 0.02)\times 10^{-3}$ for \ac{mg} with \textit{static} initial guess, and $m_\text{MG-tp} = (1.57 \pm 0.08)\times 10^{-3}$ for \ac{mg} with the \textit{time–propagated} one. 
The relative speedups in the tested range of system sizes are then given by
\[
\frac{m_\text{MG-s}}{m_\text{PM}} = 3.36 \pm 0.21 
\quad \text{and} \quad 
\frac{m_\text{MG-tp}}{m_\text{PM}} = 1.88 \pm 0.15,
\]

As a final consideration, note that while we have provided data only for the case of a molten salt system, experience with \ac{maze} calculations in other set ups (e.g., in \textit{first principles} simulations~\cite{coretti:2022}) indicates that the liquid state poses the strongest challenge to our approach. We support this statement with the data reported in Table~\ref{tab:convergence_solid_molten}, which compares the number of iterations required to reach convergence for a system of $512$ atoms simulated first as a solid at 300~K and then melted at $2800$~K, while keeping all simulation parameters identical (time step, box size, number of grid points, etc.). The last column reports the slowdown factor, defined as the ratio between the number of iterations needed for the molten salt and the corresponding solid. 
The results show that the molten phase requires more iterations to converge, reflecting the increased disorder and stronger fluctuations of the system. It is worth noting that we do not report the wall-clock time to convergence, since this quantity depends on the specific implementation of the \acl{mg} cycle (e.g., number of smoothing steps, depth of the v-cycle, and grid hierarchy), which is identical across the compared systems. The differences highlighted here therefore focus on the effect of the different phase on the convergence rate.
 
\begin{table}[ht]
    \centering
    \scriptsize
    \begin{ruledtabular}
    \begin{tabular}{lccc}
    Method & Solid & Molten salt & Slowdown factor \\
    \hline
    Multigrid - static & 7 & 22 & $\sim 2.9\times$ \\
    Multigrid - time-propagated & 3 & 11 & $\sim 3.7\times$ \\
    \ac{pmaze} & 2 & 5 & $\sim1.8\times$ \\
    \end{tabular}
    \caption{Average number of iterations required to reach a tolerance of $10^{-7}$ for solid and molten NaCl, with $\mathcal{N}=120^3$ grid points.
    The slowdown factor of the molten phase relative to the solid is reported in the last column.
    }
    \label{tab:convergence_solid_molten}
    \end{ruledtabular}
\end{table}

\subsection{Conserved quantities and properties}
In this section we analyse further the conserved quantities and properties of the algorithm.

\subsubsection{Energy}
\label{sec:conservation_energy}
\begin{figure}[htbp]
    \centering
    \includegraphics[width=0.95\linewidth]{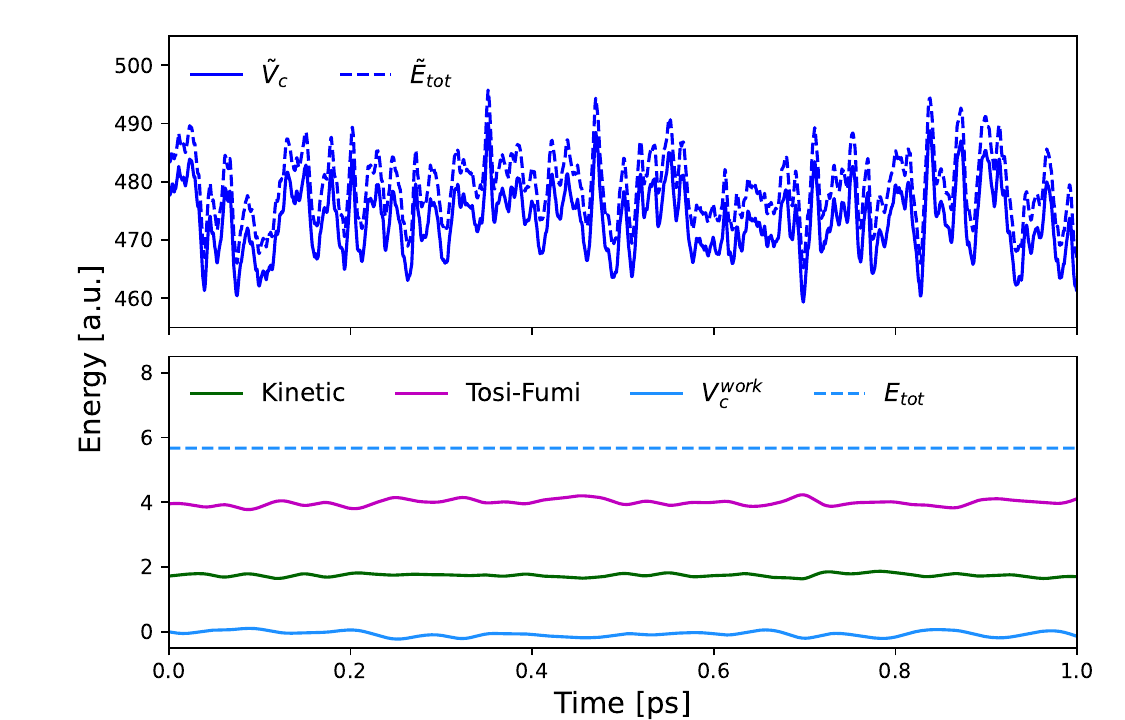}
\caption{Comparison of electrostatic and total energies. The top plot illustrates the definition given in Equation \ref{eq:energy_discr}. Notably, energy conservation is poor, as the potential energy dominates the total energy. Moreover, this energy is two orders of magnitude larger than our definition (Equation \ref{eq:E_nostra}), shown in the bottom plot, where conservation is clearly maintained. For completeness, we also show the kinetic energy and the Tosi-Fumi contributions, noting that all contributions have the same order of magnitude.}
    \label{fig:elec_new_old}
\end{figure}

Given the discussion in Section \ref{subsec:energy_force_theory},  energy conservation for the \ac{pmaze} simulation of a realistic model of a condensed system is worth discussing in some detail. In the top panel of Figure \ref{fig:elec_new_old}, we show the behaviour of Equation \ref{eq:energy_discr} for a simulation of $1$ ps of NaCl. The bottom panel of the Figure shows results for our definition of the electrostatic energy (Equation \ref{eq:E_nostra}). In the simulation, the grid spacing is $h = 0.172\,\text{\AA}$.

The figure shows that Equation~\ref{eq:energy_discr} (top panel, continuous blue curve) presents large oscillations, induced by the self-energy contributions. Furthermore, these contributions dominate the magnitude of the energy. The lower panel of the figure demonstrates the much improved conservation of the energy defined in Equation~\ref{eq:E_nostra}, and that, with this definition, potential and kinetic energies have comparable magnitude. A more precise assessment can be performed by evaluating the relative fluctuations of the total energy ($\Delta E / \langle E\rangle$) and the ratio between the standard deviation of the total energy and that of the electrostatic energy ($\Delta E / \Delta V_c$).
Using the definition from Equation \ref{eq:energy_discr}, we obtain a relative error of $\Delta \tilde{E}_\text{tot} / \langle\tilde{E}_\text{tot}\rangle = 1.3 \times 10^{-2}$, while $\Delta \tilde{E}_\text{tot} / \Delta \tilde{V}_c = 1.0$. These results confirm that electrostatic energy accounts for nearly the entire system energy. In contrast, with the definition from Equation \ref{eq:E_nostra}, we measure the significantly lower values: $\Delta E_\text{tot} / \langle E_\text{tot} \rangle = 3.2 \times 10^{-7}$ and $\Delta E_\text{tot} / \Delta V^\text{work}_c = 2.2 \times 10^{-5}$, reflecting a more stable energy representation. 

\begin{figure}[htbp]
    \centering
        \includegraphics[width=0.95\linewidth]{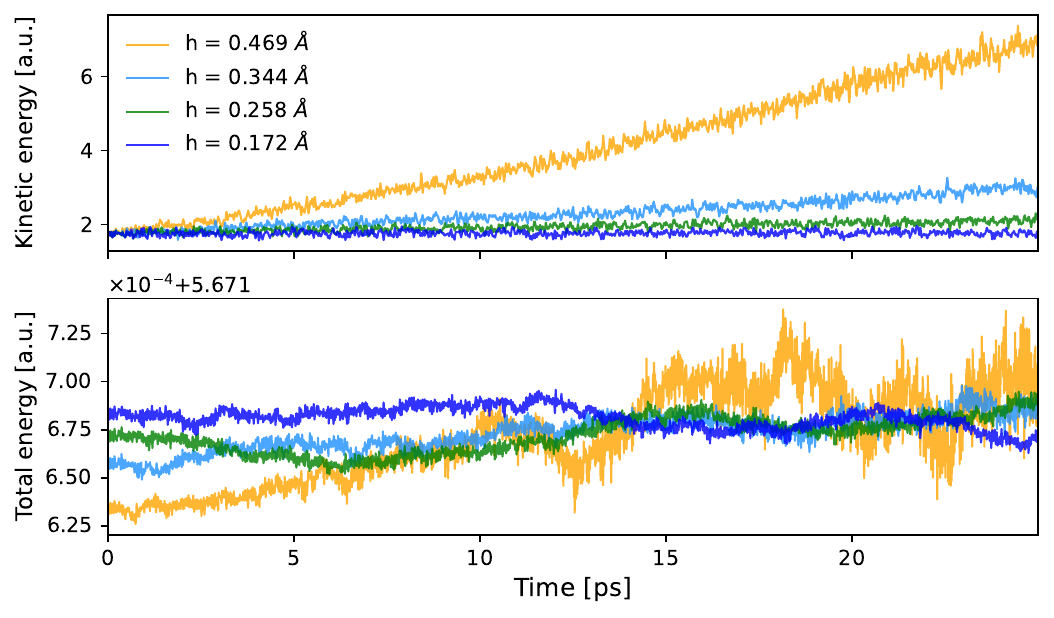}
    \caption{Time evolution of kinetic energy (top panel), total energy (middle panel). While the total energy remains well-conserved across grid resolutions, the kinetic energy shows increasing drift with coarser grids. 
    }
    \label{fig:energy_drift}
\end{figure}

\begin{figure}[htbp]
    \centering
    \includegraphics[width=0.95\linewidth]{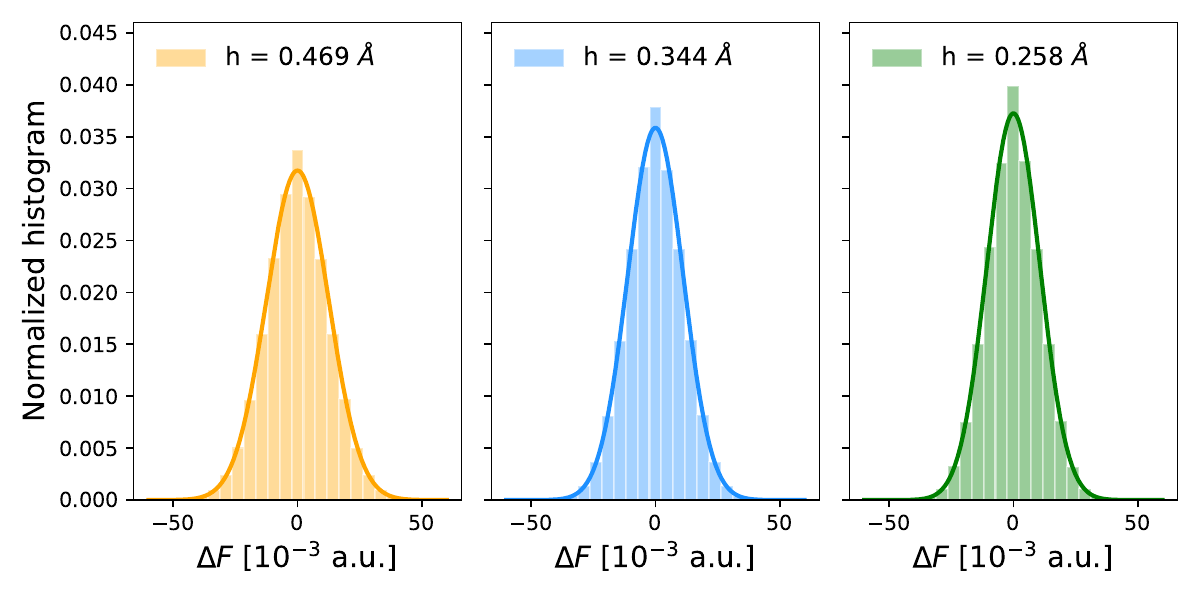}
    \caption{Normalized histograms of the mean force error for different grid spacings $h$. For each particle and at each simulation snapshot, the force deviation is computed by comparing the force obtained at a given grid spacing with a high-resolution reference computed at \( h = 0.172 \ \text{\AA} \). The distributions are centered around zero and become narrower as the resolution increases, indicating improved force accuracy.}
    \label{fig:force_spread}
\end{figure}

Next, we examine the impact of the grid spacing, $h$, on the conservation of energy as defined via Equation \ref{eq:E_nostra}. Figure~\ref{fig:energy_drift} shows the kinetic energy (top panel) and the total energy (lower panel) computed for different $h$ values.
The total energy remains well conserved across all grid spacings, even though the amplitude of its fluctuations increases as the grid becomes coarser. In contrast, the kinetic energy can display a noticeable drift over time, which becomes more pronounced with increasing $h$. This drift corresponds to a temperature instability that is negligible for finer grids but grows progressively as the grid becomes coarser. This effect probably originates from energy transfers due to inaccurate calculation of the forces at larger grid sizes. The force accuracy can be assessed by analyzing how the resolution of the computed forces varies with the grid spacing $h$. We define the reference force acting on particle  $\alpha$ as $\bm{F}_{\alpha, \text{ref}} = \bm{F}_\alpha(h=0.172 \ \text{\AA})$, corresponding to the highest-resolution simulation in this study. Then, we can define a force deviation, computed for each particle $\alpha$ and at each snapshot $t$, as:
\begin{equation}
\Delta F_\alpha(t) = \frac{1}{3} \sum_{\gamma = x, y, z} \left( F_\alpha^\gamma(t) - F_{\alpha,\text{ref}}^\gamma(t) \right)
\end{equation}
The histogram of $\Delta F_\alpha(t)$ for different values of $h$ is shown in Figure~\ref{fig:force_spread}. As the grid spacing decreases, the distributions become narrower, indicating a systematic reduction in the force error and confirming the improved accuracy of the method at higher resolutions. The time dependence and the particle index is not shown explicitly, as the deviations are aggregated over all particles and all simulation snapshots.

\subsubsection{Total momentum}
As mentioned in Section~\ref{subsec:energy_force_theory}, the definition of the force adopted in this work ensures that the total momentum of the system is conserved by the dynamics. This provides us with another control quantity to monitor along the trajectory. Numerical results show, in particular, that the level of conservation of the total momentum depends on the accuracy set for the solution of the constraints, i.e. the tolerance of the \ac{mg} solver of SHAKE. 

\begin{figure}[htbp]
    \centering
    \includegraphics[width=0.95\linewidth]{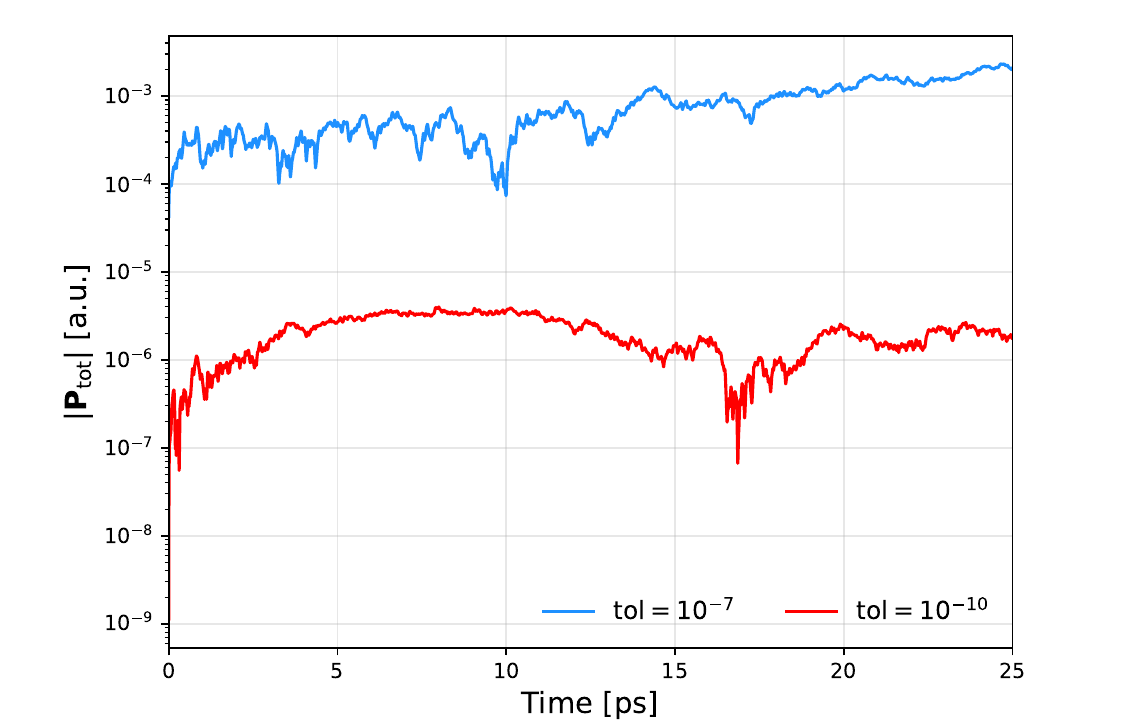}
    \caption{Conservation of total momentum for two different values of the tolerance set for \ac{mg} (tol = $10^{-7}, 10^{-10}$). The average total momentum is approximately two orders of magnitude larger than the imposed tolerance.}
    \label{fig:momentum_tot_tol}
\end{figure}

Figure \ref{fig:momentum_tot_tol} compares the conservation of the modulus of the average total momentum over time in two production runs performed with different convergence tolerances $\text{tol}=10^{-7}$ and $\text{tol}=10^{-10}$. For the looser tolerance ($10^{-7}$), the magnitude of the total momentum starts around $10^{-4}$ and shows a slight upward trend. In contrast, the stricter tolerance ($10^{-10}$) keeps the total momentum significantly lower and fluctuating around $10^{-6}$. Importantly, even with the looser tolerance of $10^{-7}$ the accuracy is sufficient to reproduce all structural and dynamical observables reported in this work.

These results highlight how the convergence tolerance impacts the accuracy of the electrostatic potential computed on the grid, and consequently the precision of the force calculation. A looser tolerance results in less accurate forces and thus larger deviations from perfect momentum conservation. Nonetheless, even in the less stringent case, the violation of total momentum conservation remains relatively small. The small non-zero value of the total momentum could be a contributing factor to the temperature drift discussed earlier. This will be further investigated in future work, and can be corrected adopting \textit{ad hoc} subtractions schemes as suggested in the literature\cite{Rottler:2004} or imposing strict momentum conservation as an additional constraint.

\subsubsection{Time reversibility}
\begin{figure}[ht!]
\centering
\includegraphics[width=0.95\linewidth]{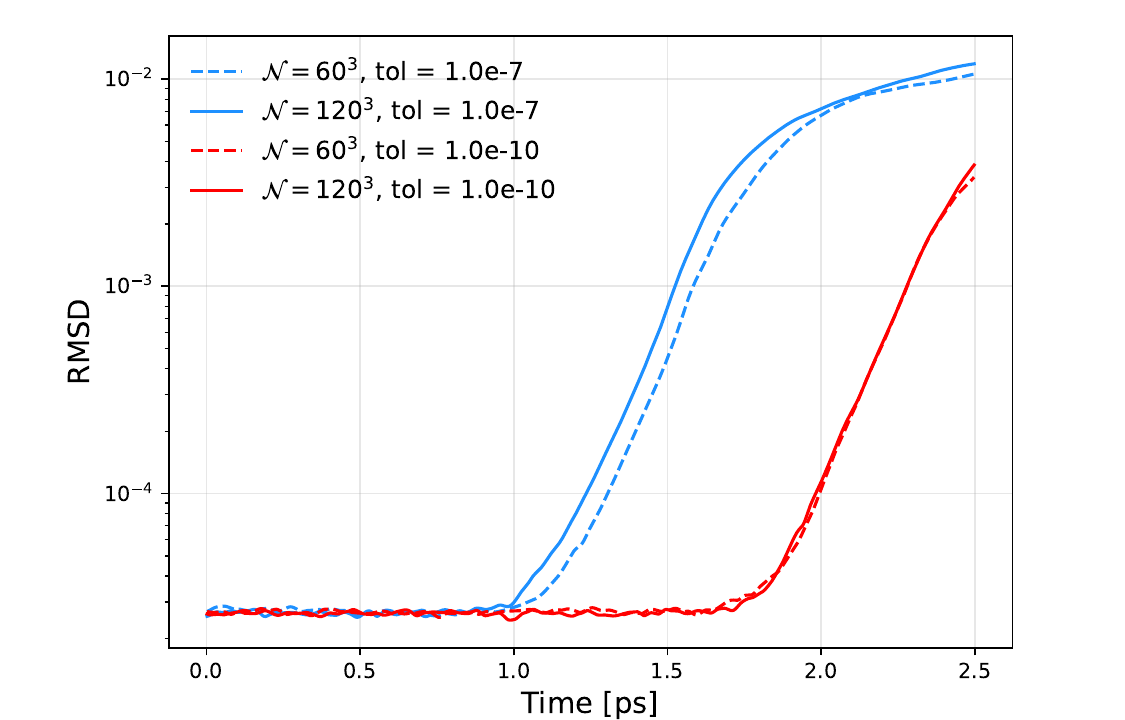}
\caption{Root mean square deviation (RMSD) between the forward and backward trajectories as a function of time, normalized by the number of particles and the box side length $L$. Results are shown for different values of the number of grid points $\mathcal{N}$ and \ac{mg} tolerances.}
\label{fig:time_rev_rmsd}
\end{figure}

Time reversibility, and indeed symplecticity, of \ac{pmaze} is formally guaranteed, for sufficiently accurate solution of the constraints, by the combination of time reversible algorithms for the evolution of the physical and auxiliary degrees of freedom and SHAKE~\cite{leimkuhler1999comparison}. Monitoring this property then provides additional information on the tolerance needed to reliably propagate the system. To assess the time reversal behavior of the system, we consider the root mean square deviation (RMSD) between a forward propagated trajectory and its backward counterpart obtained after inversion of the sign of momenta of the particles. The RMSD is normalized by the number of particles and the box side length $L$:
\begin{equation}
\text{RMSD}(t) = \frac{1}{N_p L}\sum_{\alpha=1}^{N_p}\sqrt{[\bm{r}_{\alpha}^\text{backward}(\mathcal{T} - t) - \bm{r}_{\alpha}^\text{forward}(t)]^2}
\end{equation}
The quantity above was computed in two simulations of the NaCl system, with values of the \acl{mg}'s tolerance set to $10^{-7}$ and $10^{-10}$, respectively. In these simulations, standard forward simulations of duration $\mathcal{T} = 2.5$ ps were carried out, followed by backward simulations of equal duration. In Figure~\ref{fig:time_rev_rmsd}, we plot  $\text{RMSD}(t)$ for tolerance $10^{-7}$ in blue and in red for the $10^{-10}$. To investigate the effect of the accuracy of the discretization, we also report results for different number of grid points: the dashed curves have been obtained with $\mathcal{N}=60^3$, while the continuous curves correspond to $\mathcal{N}=120^3$. The figure illustrates that time reversibility is better preserved (for this simulation length) with the more stringent threshold, while the onset of an exponential violation is observed for $\text{tol}=10^{-7}$ already after $1$ ps. We note that deviations from strict time reversibility are unavoidable at longer times due to numerical inaccuracies in the integration of the evolution equations.
Interestingly, the grid spacing has a minor effect as demonstrated by the close agreement of the results for different grid sizes. 

\subsubsection{Stationarity}
\begin{figure}[htbp]
    \centering
    \includegraphics[width=0.95\linewidth]{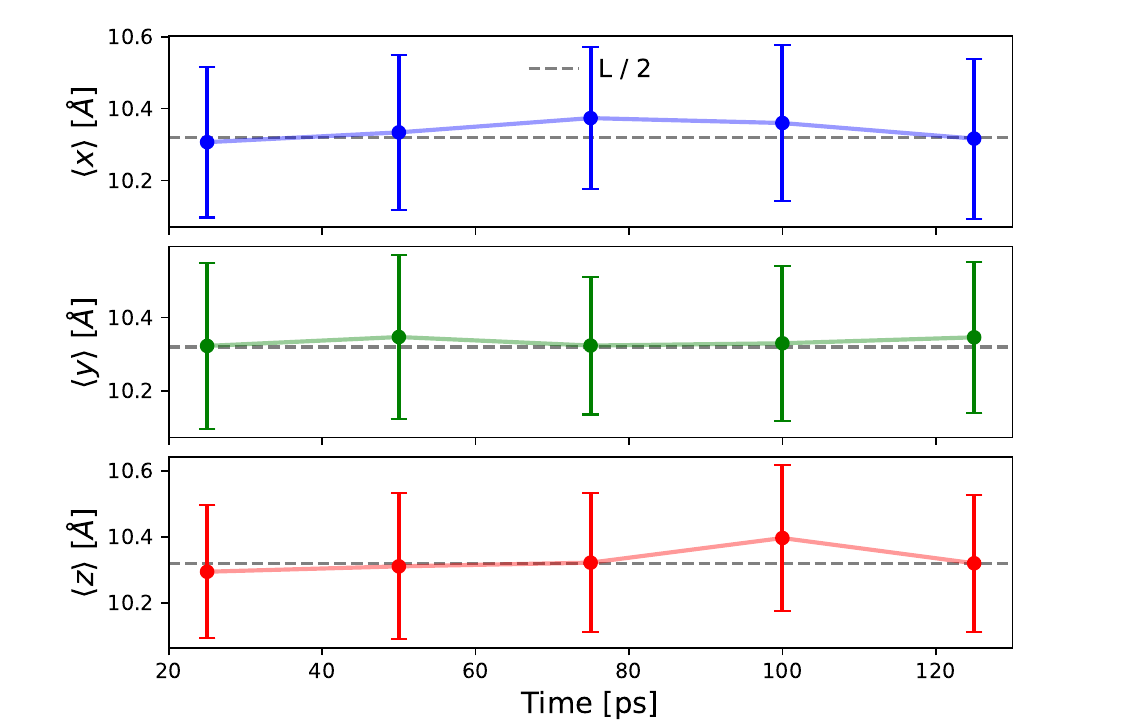}
    \caption{Average particle positions over time, grouped into five 25 ps intervals. The values remain stable around $L/2$, confirming simulation stationarity.}
    \label{fig:stationarity}
\end{figure}

Another property of interest is the stationarity of the simulation, and in particular the independence of the value of time averages on the time origin of the average. To investigate this property for our algorithm, we performed a run of duration $\mathcal{T} = 125$ ps using the particles' position as our test quantity. 
We first computed the average (over the number of particles) of the position at each step of the simulation. Then, the array of the average positions was divided into five slots of $25$ ps each. For each lag, the average with respect to time of the position and its standard deviation for each Cartesian component was calculated and plotted as a function of time, as shown in Figure \ref{fig:stationarity}.
The measured averages are equal to each other within error bars and - as expected - equal to half the size of the simulation box ($L/2$) confirming the stationarity of the simulation.

 \section{Conclusions}
In this work, we introduced \ac{pmaze}, a novel method for computing electrostatic forces in molecular dynamics simulations. 
The method belongs to the class of real-space solvers, offering an alternative to reciprocal-space techniques. It differs from previous real-space solvers in that it does not tackle the Poisson equation directly, but via an extended dynamical system in which the values of the discretized electrostatic potential are treated as auxiliary degrees of freedom that are constrained to satisfy the Poisson equation. The constraint is imposed by solving a linear system of equations that has numerical advantages over the one appearing in previous approaches. Through a series of validation tests, we showed that \ac{pmaze} provides correct electrostatic potentials and forces, both in proof-of-concept simulations (Appendix A) and for a realistic model of a condensed phase system.  Additionally, the method preserves work-derived energy and total momentum, maintains time reversibility, and ensures stationarity throughout the simulation. Previous theoretical work~\cite{bonella:2020} proved that \ac{maze} approaches sample the exact probability density for the physical \ac{dof}. \\
From a performance perspective, \ac{pmaze} inherits the asymptotic scaling of the linear solver chosen to solve the constraint condition.  This was demonstrated considering \acl{mg} and conjugate gradient (see Appendix~\ref{appendix_B:cg}). This flexibility means that any optimized solver can be embedded in the \ac{pmaze} framework without altering its structure or algorithm. In particular, the approach enjoys linear scaling with the number of particles when adopting the \ac{mg} solver for the constraints.  Furthermore, one of the challenges for real-space solvers lies in the number of iterations required for convergence. By reformulating the problem, \ac{pmaze} reduces this number substantially compared to direct solutions of the Poisson equation, accelerating convergence. Future work will focus on verifying these very encouraging results on a broader range of systems.

\begin{acknowledgments}
The authors are grateful to B. Rotenberg for insightful discussions. 

This research was supported by the NCCR MARVEL, a National Centre for Competence in Research, funded by the Swiss National Science Foundation (grant number 205602). J.N. acknowledges funding via the NCCR MARVEL INSPIRE Potential master fellowship that supported her stay at the Centre Europ\'een de Calcul Atomique et Mol\'eculaire (CECAM). 
The work was supported also by the National Institute of Health (NIH) via grant R35-GM152124 (B.R.).
\end{acknowledgments}

The authors have no conflicts to disclose.

\begin{appendix}

\section{Two-particle model}
\label{sec:Appendix1}
In this Appendix, we consider a minimal two-particle model to validate the ability of \ac{pmaze} to solve the discrete Poisson equation and to examine the behavior of different energy and force estimators. We first demonstrate that \ac{pmaze} correctly reproduces the electrostatic potential for two point-like charges. Then, we show the merits of Equation \ref{eq:F_nostra} over the alternative introduced at the end of Section~\ref{subsec:energy_force_theory}. We shall also discuss the behaviour of different energy estimators. In the model, the two particles are positioned along the x-axis and PBCs are enforced. To benchmark accuracy, we compare the value of the potential computed with \ac{pmaze} with the result obtained using the iterative method known as the \ac{sor} method\cite{Klapper:1986,Nicholls:1991} as coded in CHARMM (version c47b1) in the PBEQ package\cite{Im:1998,CHARMM-GUI:2008}. We use \ac{sor} as our benchmark because, even for a simple system of two charges, no analytical solution exists for the electrostatic potential under PBC. Nevertheless, the solution of our problem approaches the Coulomb potential near the charges, provided they are point-like and sufficiently far from the boundaries. 

\begin{figure}[htbp]
    \centering
    \includegraphics[width=0.95\linewidth]{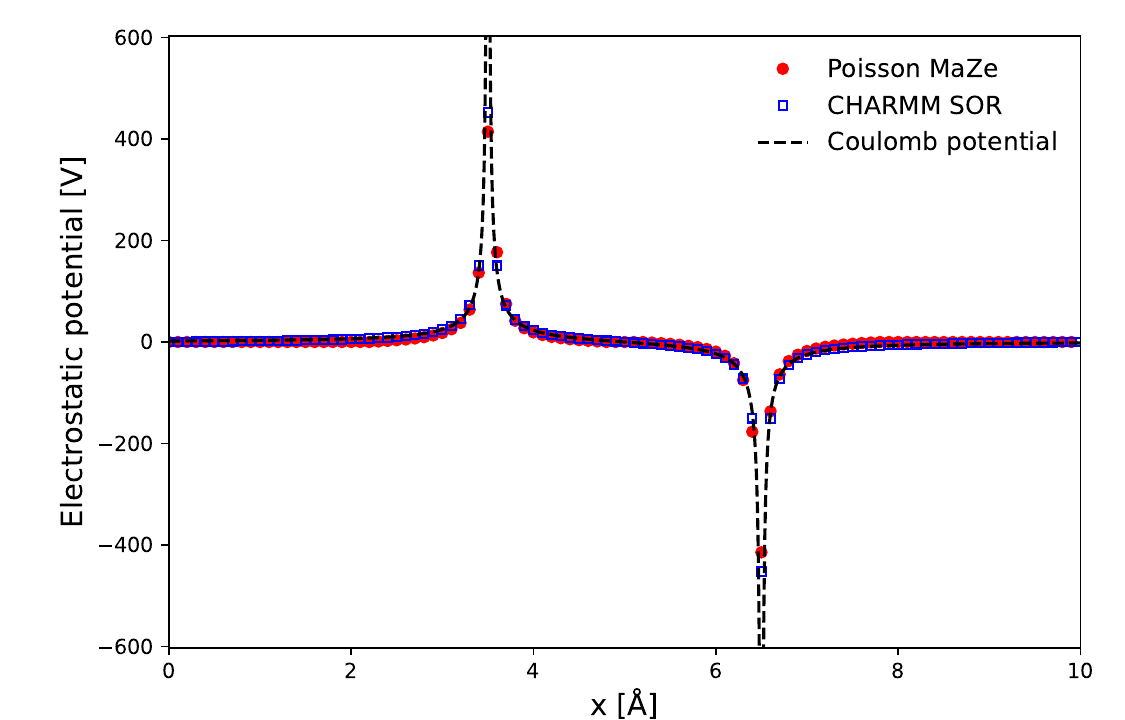}
    \caption{Section of the electrostatic potential generated by two particles with opposite charges plotted along the x-axis. The figure shows the overlap between the potential computed with \ac{pmaze} (in red), with a grid of $\mathcal{N}=10^6$ points, and CHARMM \ac{sor} (in blue). The results of the two algorithms are in excellent agreement and match closely the Coulomb potential (dashed grey line).}
    \label{fig:elec_potential}
\end{figure}

Figure \ref{fig:elec_potential} demonstrates that the results obtained with \ac{pmaze} (red dots) and CHARMM \ac{sor} (blue squares) agree very well, and that both approach the Coulomb potential (drawn as the dashed line) for the chosen set up.
\noindent
We now illustrate the behavior of the energy estimators defined in Equation~\ref{eq:E_nostra} and Equation~\ref{eq:energy_discr}, with the goal of highlighting the pathology induced by self-term. The same calculation will also enable us to further motivate the choice of Equation~\ref{eq:F_nostra} as an estimator of the force.  Let us consider a one-dimensional simulation, where two charged particles are placed in a box of size $L = 10 \ \text{\AA}$. Particle $A$, positively charged, is placed at $x_A = 5 \ \text{\AA}$, while the initial position of Particle $B$ is $x_B = 7 \ \text{\AA}$. Then, Particle $B$ is manually displaced with a fixed displacement $\Delta x = 0.1 \ \text{\AA}$ until it returns to its initial position. For every displacement, we measure the electrostatic energy of the system and the force acting on Particle $B$ with our different estimators. To isolate the effect of the self-energy we need a procedure to measure it. This can be achieved by considering a simulation box in which only one of the particles, say Particle \( B \), is present. To ensure overall charge neutrality of the simulation box, we use a compensating uniform background charge. The self-energy of particle \( B \) is computed by evaluating the electrostatic energy of this system. Note that this procedure to compute the self-energy cannot be generalized to highly dimensional, interacting, systems.

\begin{figure}[htbp]
\centering
  \includegraphics[width=0.95\linewidth]{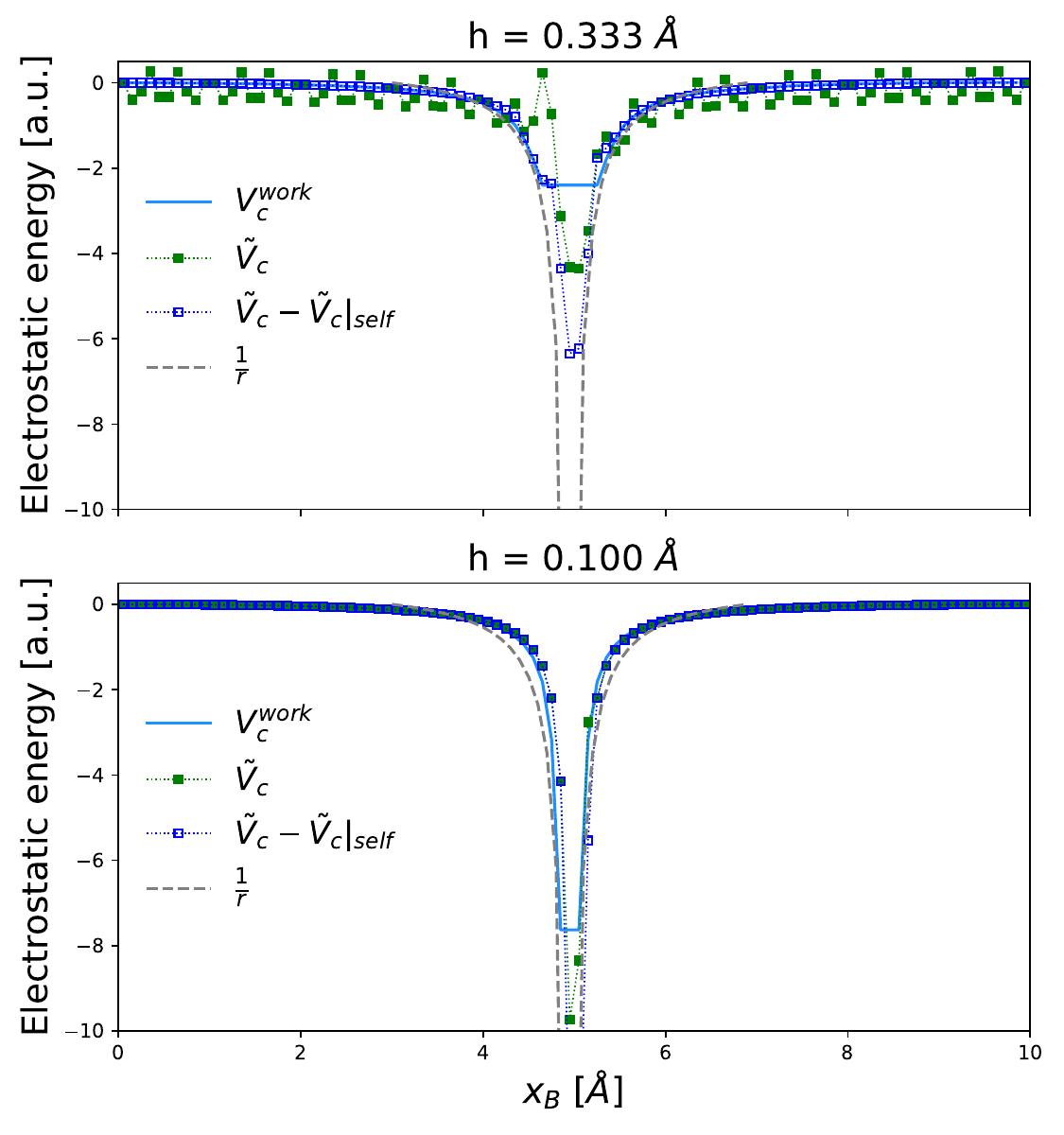}
\caption{Comparison between the different expressions for the energies in 1D for $h = 0.333 \ \text{\AA}$ (upper panel) and $h = 0.1 \ \text{\AA}$ (lower panel). The electrostatic energy computed using the work is shown in blue. Oscillations of the discretized expression for the energy decrease with decreasing $h$.}
\label{fig:energy_comparison1D}
\end{figure}

In Figure~\ref{fig:energy_comparison1D}, we show the comparison of the energy estimators. The upper panel of the figure reports results with a (relatively coarse) grid spacing of $h = 0.333\ \text{\AA}$ (corresponding to $\mathcal{N} = 30^3$), while the lower panel considers a finer grid with $h = 0.1 \ \text{\AA}$ ($\mathcal{N} = 10^6$). In the figure, the green line with filled squares corresponds to estimator $\tilde{V}_c$ defined in Equation \ref{eq:energy_discr}. The corresponding curve in the upper panel exhibits noticeable fluctuations that originate from the self-energy contribution. This is demonstrated by considering the blue curve with open squares that reports values of the same estimator after subtraction of the calculated self-energy contributions. These values agree with the results obtained using Equation~\ref{eq:E_nostra} and are shown in the figure as the blue curve. Note that this curve is not affected by oscillations originating from the self-energy. It can be observed that the value of this energy estimator remains constant when the distance between Particle $A$ and Particle $B$ is smaller than the grid size $h$. In this case, the particles share the same nearest neighbors, resulting in a total charge of zero, which prevents any increase in the work. However, this nonphysical scenario would not occur in dynamical simulations due to the presence of a repulsive potential between the charges. The results in the lower panel of Figure \ref{fig:energy_comparison1D} indicate that, with the chosen charge assignment scheme, the self-energy fluctuations decrease with decreasing grid spacing $h$, leading to excellent agreement between the two estimators of the energy. 

\begin{figure}[htbp]
\centering
  \includegraphics[width=0.95\linewidth]{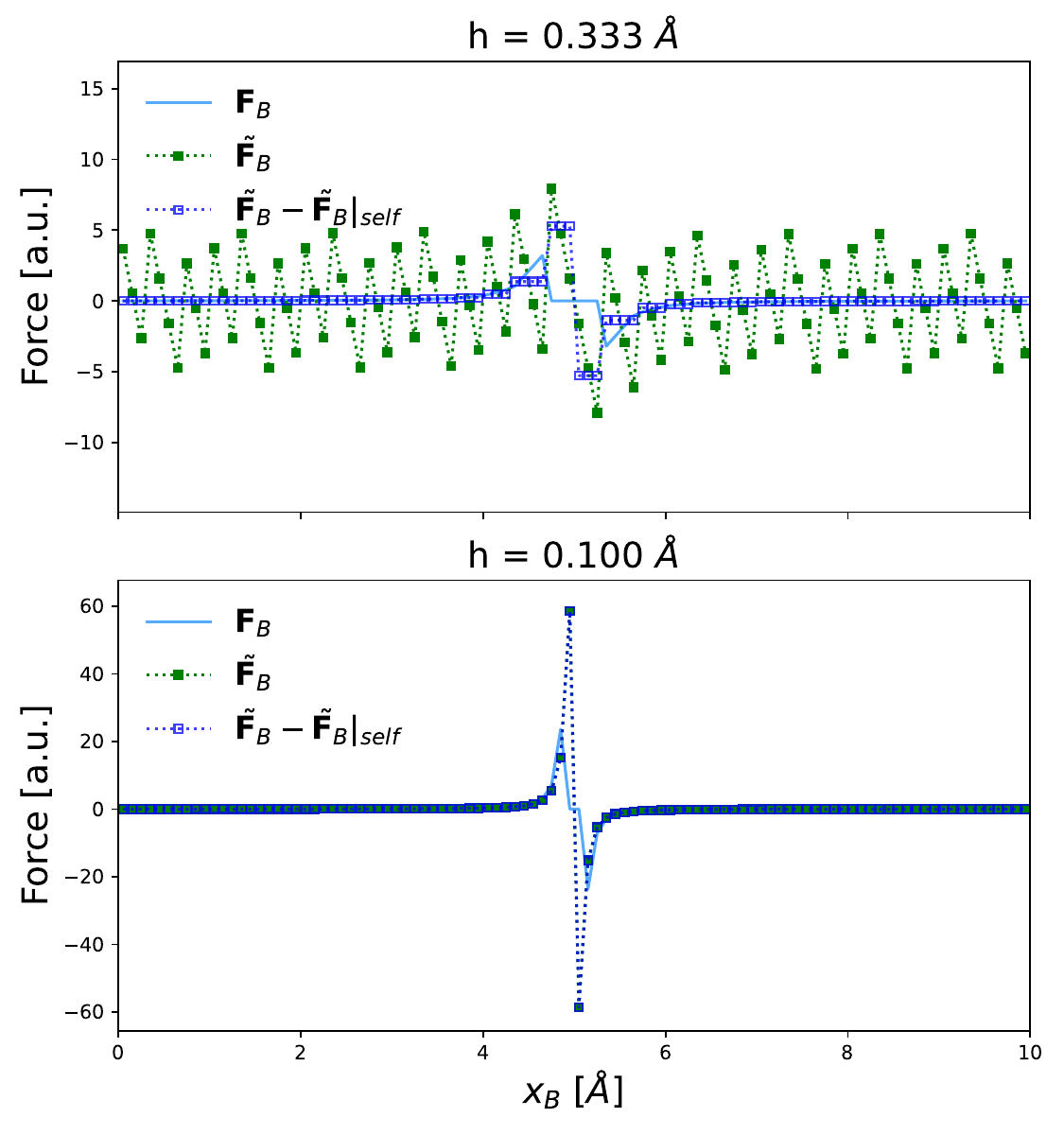}
\caption{Comparison between the different expressions for the forces acting on Particle $B$ in 1D for $h = 0.333 \ \text{\AA}$ (upper panel) and $h = 0.1 \ \text{\AA}$ (lower panel). $\bm{\tilde{F}}_B$ matches overall the difference between the force $\bm{F}_B$ and the self-force contribution. Some discrepancies around the positions of the particles arise from the charge distribution along the neighbors. For $h = 0.1 \ \text{\AA}$ the results match almost exactly. The larger vertical scale for $h = 0.1 \ \text{\AA}$ reflects the finer resolution of the force singularity at short distances between the particles.}
\label{fig:force_comparison1D}
\end{figure}

Figure \ref{fig:force_comparison1D} shows the force acting on particle $B$ computed via Equation~\ref{eq:F_nostra}, blue curve and noted as $\bm{F}_B$, and Equation~\ref{eq:F_Benoit}, green filled squares and noted as $\bm{\tilde{F}}_B$. The upper panel, reporting results with the coarse grid spacing $h = 0.333 \ \text{\AA}$, shows the effect of the self-energy oscillations on $\bm{\tilde{F}}_B$, and -- as for the energy estimators -- these are removed by subtracting the self-energy contribution (empty blue symbols). After this correction, $\bm{\tilde{F}}_B$ matches well $\bm{F_B}$, with the exception of a small region close to $A$. The discrepancies can be explained by the assignment of the charge distribution to neighboring points in the grid. As shown in the lower panel of the figure, corresponding to $h = 0.1 \ \text{\AA}$, $\bm{\tilde{F}}_B$ oscillations essentially disappear as the grid spacing decreases. As for the energy, $\bm{F}_B$ does not show oscillations for any grid size.

\section{Performance and scaling of \ac{pmaze} using conjugate gradient method}
\label{appendix_B:cg}
In this Appendix, we analyze the performance and scaling of \ac{pmaze} when used in combination with the \ac{plcg}\cite{coretti:2022,Hestenes:1952, elber:2011}. We implemented the method with a simple Jacobi preconditioner, with elements $J_{mn}=M_{mn}\delta_{mn}$. This study shows that the acceleration provided by the \ac{maze} formulation is not solver-specific but stems from an improved conditioning of the problem. Comparing \ac{plcg} with and without the \ac{maze} superstructure confirms that \ac{pmaze} reduces the iterations needed for convergence, demonstrating its benefits beyond the \acl{mg} case discussed in the main text. 

\begin{figure}[ht!]
\centering
\includegraphics[width=0.95\linewidth]{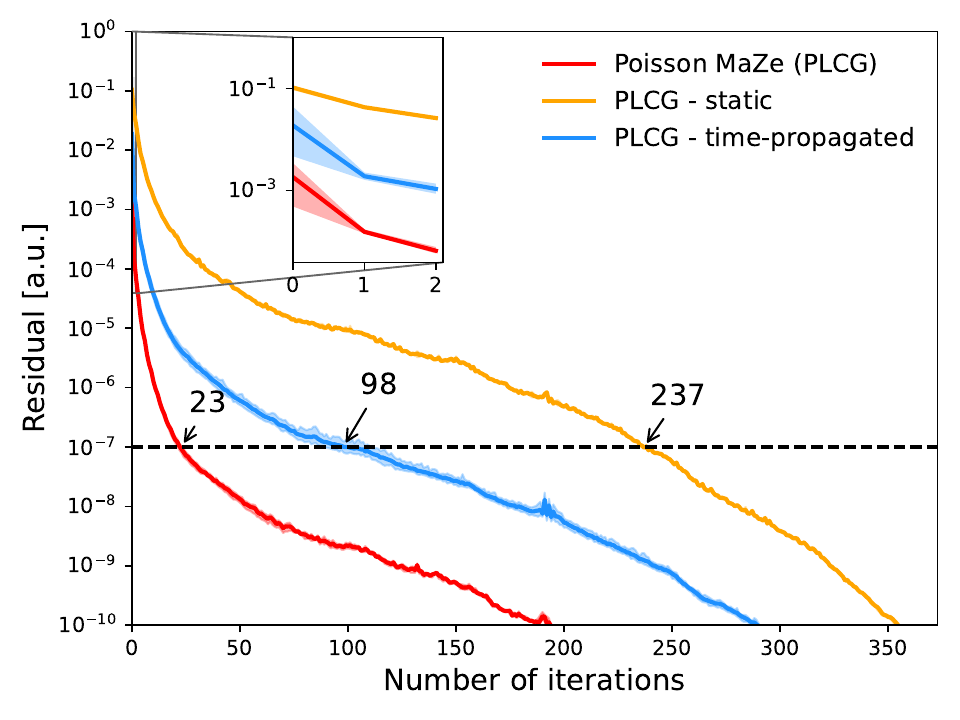}
\caption{Convergence of \ac{pmaze} using \ac{plcg} (red), \ac{plcg} with \textit{static} (yellow), 
and \textit{time-propagated} (blue) starting value for the iterations. Solid lines show the mean residual across trajectories, and shaded bands indicate the pointwise min–max envelope. Arrows mark the iteration at which the residual crosses $10^{-7}$.}
\label{fig:residual_cg}
\end{figure}

In Figure~\ref{fig:residual_cg}, we present the same analysis presented in Section~\ref{subsec:convergence} for the \ac{mg} method. In the figure we report residuals as a function of iterations, with solid lines indicating averages and shaded bands showing variability; the \ac{pmaze} curve (red) converges faster than both \textit{static} and \textit{time–propagated} \ac{plcg}.
For a tolerance of $10^{-7}$, \ac{pmaze} converges in only $23$ iterations (see arrows in Figure~\ref{fig:tol_vs_v_cycles}), whereas the \textit{time-propagated} \ac{plcg} scheme requires roughly four times as many iterations, and the \textit{static} \ac{plcg} roughly ten times as many. 
The inset of Figure~\ref{fig:tol_vs_v_cycles} shows the effect of the different choices for the starting value of the iterations, confirming that - as for the case discussed in the main text - the combined use of the putative potential and of the previous value of the constraints, leads to a better starting value for \ac{pmaze}.

We now discuss the scaling of \ac{pmaze} with respect to both the grid size and the number of physical particles in the system. The number of iterations is determined by imposing a stopping tolerance $\rm{tol}=10^{-10}$.
Let us consider first the scaling of the algorithm with the number of grid points $\mathcal{N}$. 
Through the convergence analysis of iterative methods (see, e.g., \cite{polyak:1987}) and the operation count, we can predict that the number of iterations scales approximately as $\mathcal{O}(\mathcal{N}^{1/3})$, while the time per iteration scales approximately as $\mathcal{O}(\mathcal{N})$. The iteration count is expected to grow proportionally to ($\sqrt{\kappa}$), where ($\kappa$) is the condition number of the system matrix. For the Laplacian operator, the condition number scales as $\kappa \sim \frac{\lambda_{\text{max}}}{\lambda_{\text{min}}} \sim N^2$ ($\lambda_\text{max,min}$ being the biggest and smallest eigenvalues), where \(N\) is the linear grid size. Thus, we expect it to scale as $\sqrt{\kappa} \sim N \sim \mathcal{N}^{1/3}$.
Moreover, the cost of each iteration is of order $\mathcal{O}(\mathcal{N})$, since the most expensive operation in the \ac{plcg} algorithm is the matrix-vector product. The overall scaling of the algorithm as a function of the number of grid points is given by the product of the two contributions, resulting in $\mathcal{O}(\mathcal{N}^{4/3})$.
Since the \ac{plcg} is sensitive to the form of the right–hand side \cite{Beckermann:2002}, we performed six simulations to investigate this effect, varying only the number of particles $N_p$ while keeping the simulation box size ($L=37.15 \ \text{\AA}$) and the number of grid points ($\mathcal{N} = 180^3$) fixed. 

\begin{figure}[htbp]
\centering
\includegraphics[width=0.95\linewidth]{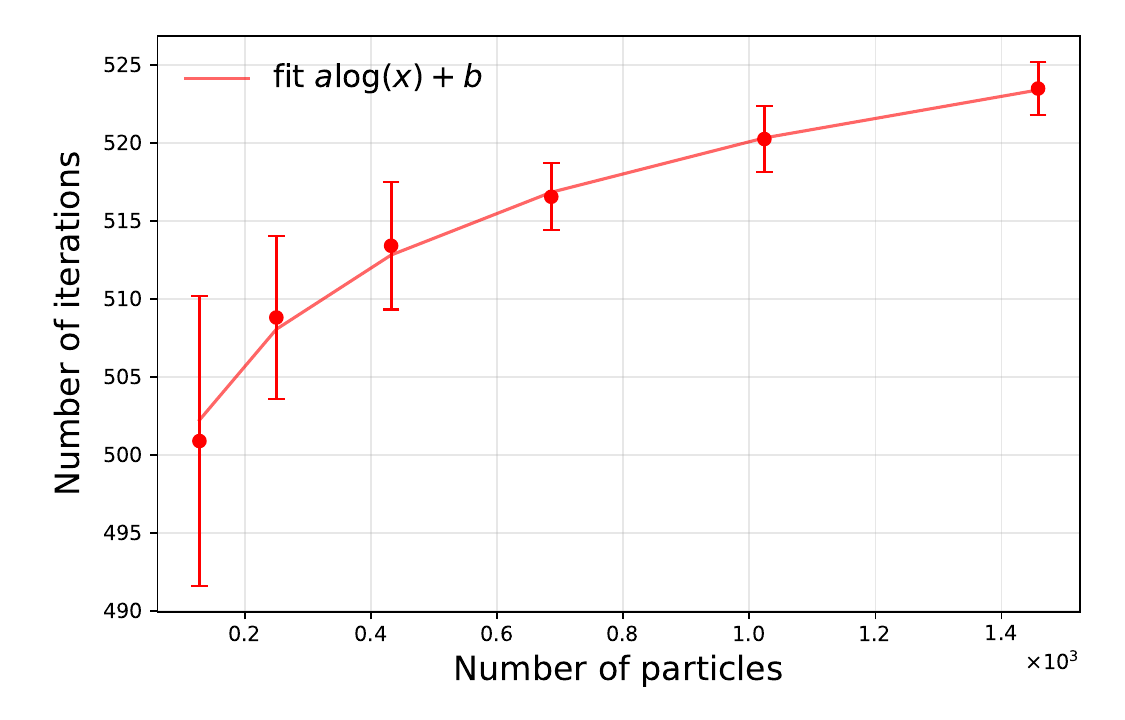}
\caption{Scaling analysis of \ac{pmaze} as a function of the number of particles, showing the number of iterations required for \ac{plcg} convergence.}
\label{fig:scaling_particles}
\end{figure}

In the following we varied $N_p$ in the range $N_p=128$ to $N_p=1458$. Equilibration and data productions were performed as described in Section \ref{subsec:performance}. Figure 
\ref{fig:scaling_particles} shows the scaling of the number of iterations needed for convergence, which follows approximately $\mathcal{O}(\log N_p)$, reflecting the dependence on the right-hand side.
Our numerical tests showed also that the time per iteration is essentially independent on $N_p$.  
The results reported in this section indicate that the overall scaling of \ac{pmaze} with a \ac{plcg} solver is $\mathcal{O}(\mathcal{N}^{4/3} \log N_p)$. Including the standard cost of calculating electrostatic and non-electrostatic forces, which can be reduced to linear in the number of particles, the overall cost of the algorithm scales as  $\mathcal{O}(\mathcal{N}^{4/3} \log N_p + N_p)$. Typically, however, scaling with respect to the number of grid points is the dominant contribution. 

\end{appendix}

\bibliography{main}

\begin{thebibliography}{54}%
\makeatletter
\providecommand \@ifxundefined [1]{%
 \@ifx{#1\undefined}
}%
\providecommand \@ifnum [1]{%
 \ifnum #1\expandafter \@firstoftwo
 \else \expandafter \@secondoftwo
 \fi
}%
\providecommand \@ifx [1]{%
 \ifx #1\expandafter \@firstoftwo
 \else \expandafter \@secondoftwo
 \fi
}%
\providecommand \natexlab [1]{#1}%
\providecommand \enquote  [1]{``#1''}%
\providecommand \bibnamefont  [1]{#1}%
\providecommand \bibfnamefont [1]{#1}%
\providecommand \citenamefont [1]{#1}%
\providecommand \href@noop [0]{\@secondoftwo}%
\providecommand \href [0]{\begingroup \@sanitize@url \@href}%
\providecommand \@href[1]{\@@startlink{#1}\@@href}%
\providecommand \@@href[1]{\endgroup#1\@@endlink}%
\providecommand \@sanitize@url [0]{\catcode `\\12\catcode `\$12\catcode `\&12\catcode `\#12\catcode `\^12\catcode `\_12\catcode `\%12\relax}%
\providecommand \@@startlink[1]{}%
\providecommand \@@endlink[0]{}%
\providecommand \url  [0]{\begingroup\@sanitize@url \@url }%
\providecommand \@url [1]{\endgroup\@href {#1}{\urlprefix }}%
\providecommand \urlprefix  [0]{URL }%
\providecommand \Eprint [0]{\href }%
\providecommand \doibase [0]{http://dx.doi.org/}%
\providecommand \selectlanguage [0]{\@gobble}%
\providecommand \bibinfo  [0]{\@secondoftwo}%
\providecommand \bibfield  [0]{\@secondoftwo}%
\providecommand \translation [1]{[#1]}%
\providecommand \BibitemOpen [0]{}%
\providecommand \bibitemStop [0]{}%
\providecommand \bibitemNoStop [0]{.\EOS\space}%
\providecommand \EOS [0]{\spacefactor3000\relax}%
\providecommand \BibitemShut  [1]{\csname bibitem#1\endcsname}%
\let\auto@bib@innerbib\@empty
\bibitem [{\citenamefont {Allen}\ and\ \citenamefont {Tildesley}(1989)}]{Allen-Tildesley}%
  \BibitemOpen
  \bibfield  {author} {\bibinfo {author} {\bibfnamefont {M.}~\bibnamefont {Allen}}\ and\ \bibinfo {author} {\bibfnamefont {D.}~\bibnamefont {Tildesley}},\ }\href@noop {} {\emph {\bibinfo {title} {Computer Simulation of Liquids}}}\ (\bibinfo  {publisher} {Oxford Science Publications, Clarendon Press},\ \bibinfo {address} {Oxford},\ \bibinfo {year} {1989})\BibitemShut {NoStop}%
\bibitem [{\citenamefont {Darden}, \citenamefont {York},\ and\ \citenamefont {Pedersen}(1993)}]{Darden:1993}%
  \BibitemOpen
  \bibfield  {author} {\bibinfo {author} {\bibfnamefont {T.}~\bibnamefont {Darden}}, \bibinfo {author} {\bibfnamefont {D.}~\bibnamefont {York}}, \ and\ \bibinfo {author} {\bibfnamefont {L.}~\bibnamefont {Pedersen}},\ }\href@noop {} {\bibfield  {journal} {\bibinfo  {journal} {The Journal of Chemical Physics}\ }\textbf {\bibinfo {volume} {98}},\ \bibinfo {pages} {10089} (\bibinfo {year} {1993})}\BibitemShut {NoStop}%
\bibitem [{\citenamefont {Essmann}\ \emph {et~al.}(1995)\citenamefont {Essmann}, \citenamefont {Perera}, \citenamefont {Berkowitz}, \citenamefont {Darden}, \citenamefont {Lee},\ and\ \citenamefont {Pedersen}}]{Essmann:1995}%
  \BibitemOpen
  \bibfield  {author} {\bibinfo {author} {\bibfnamefont {U.}~\bibnamefont {Essmann}}, \bibinfo {author} {\bibfnamefont {L.}~\bibnamefont {Perera}}, \bibinfo {author} {\bibfnamefont {M.~L.}\ \bibnamefont {Berkowitz}}, \bibinfo {author} {\bibfnamefont {T.}~\bibnamefont {Darden}}, \bibinfo {author} {\bibfnamefont {H.}~\bibnamefont {Lee}}, \ and\ \bibinfo {author} {\bibfnamefont {L.~G.}\ \bibnamefont {Pedersen}},\ }\href@noop {} {\bibfield  {journal} {\bibinfo  {journal} {The Journal of Chemical Physics}\ }\textbf {\bibinfo {volume} {103}},\ \bibinfo {pages} {8577} (\bibinfo {year} {1995})}\BibitemShut {NoStop}%
\bibitem [{\citenamefont {Darden}, \citenamefont {Toukmaji},\ and\ \citenamefont {Pedersen}(1997)}]{Darden-1997}%
  \BibitemOpen
  \bibfield  {author} {\bibinfo {author} {\bibfnamefont {T.}~\bibnamefont {Darden}}, \bibinfo {author} {\bibfnamefont {A.}~\bibnamefont {Toukmaji}}, \ and\ \bibinfo {author} {\bibfnamefont {L.}~\bibnamefont {Pedersen}},\ }\href@noop {} {\bibfield  {journal} {\bibinfo  {journal} {Journal de Chimie Physique et de Physico-Chimie Biologique}\ }\textbf {\bibinfo {volume} {94}},\ \bibinfo {pages} {1346} (\bibinfo {year} {1997})}\BibitemShut {NoStop}%
\bibitem [{\citenamefont {Hwang}\ \emph {et~al.}(2024)\citenamefont {Hwang}, \citenamefont {Austin}, \citenamefont {Blondel}, \citenamefont {Boittier}, \citenamefont {Boresch}, \citenamefont {Buck}, \citenamefont {Buckner}, \citenamefont {Caflisch}, \citenamefont {Chang}, \citenamefont {Cheng} \emph {et~al.}}]{hwang2024charmm}%
  \BibitemOpen
  \bibfield  {author} {\bibinfo {author} {\bibfnamefont {W.}~\bibnamefont {Hwang}}, \bibinfo {author} {\bibfnamefont {S.~L.}\ \bibnamefont {Austin}}, \bibinfo {author} {\bibfnamefont {A.}~\bibnamefont {Blondel}}, \bibinfo {author} {\bibfnamefont {E.~D.}\ \bibnamefont {Boittier}}, \bibinfo {author} {\bibfnamefont {S.}~\bibnamefont {Boresch}}, \bibinfo {author} {\bibfnamefont {M.}~\bibnamefont {Buck}}, \bibinfo {author} {\bibfnamefont {J.}~\bibnamefont {Buckner}}, \bibinfo {author} {\bibfnamefont {A.}~\bibnamefont {Caflisch}}, \bibinfo {author} {\bibfnamefont {H.-T.}\ \bibnamefont {Chang}}, \bibinfo {author} {\bibfnamefont {X.}~\bibnamefont {Cheng}},  \emph {et~al.},\ }\href@noop {} {\bibfield  {journal} {\bibinfo  {journal} {The Journal of Physical Chemistry B}\ }\textbf {\bibinfo {volume} {128}},\ \bibinfo {pages} {9976} (\bibinfo {year} {2024})}\BibitemShut {NoStop}%
\bibitem [{\citenamefont {Salomon-Ferrer}, \citenamefont {Case},\ and\ \citenamefont {Walker}(2013)}]{salomon2013overview}%
  \BibitemOpen
  \bibfield  {author} {\bibinfo {author} {\bibfnamefont {R.}~\bibnamefont {Salomon-Ferrer}}, \bibinfo {author} {\bibfnamefont {D.~A.}\ \bibnamefont {Case}}, \ and\ \bibinfo {author} {\bibfnamefont {R.~C.}\ \bibnamefont {Walker}},\ }\href@noop {} {\bibfield  {journal} {\bibinfo  {journal} {Wiley Interdisciplinary Reviews: Computational Molecular Science}\ }\textbf {\bibinfo {volume} {3}},\ \bibinfo {pages} {198} (\bibinfo {year} {2013})}\BibitemShut {NoStop}%
\bibitem [{\citenamefont {Abraham}\ \emph {et~al.}(2015)\citenamefont {Abraham}, \citenamefont {Murtola}, \citenamefont {Schulz}, \citenamefont {P{\'a}ll}, \citenamefont {Smith}, \citenamefont {Hess},\ and\ \citenamefont {Lindahl}}]{abraham2015gromacs}%
  \BibitemOpen
  \bibfield  {author} {\bibinfo {author} {\bibfnamefont {M.~J.}\ \bibnamefont {Abraham}}, \bibinfo {author} {\bibfnamefont {T.}~\bibnamefont {Murtola}}, \bibinfo {author} {\bibfnamefont {R.}~\bibnamefont {Schulz}}, \bibinfo {author} {\bibfnamefont {S.}~\bibnamefont {P{\'a}ll}}, \bibinfo {author} {\bibfnamefont {J.~C.}\ \bibnamefont {Smith}}, \bibinfo {author} {\bibfnamefont {B.}~\bibnamefont {Hess}}, \ and\ \bibinfo {author} {\bibfnamefont {E.}~\bibnamefont {Lindahl}},\ }\href@noop {} {\bibfield  {journal} {\bibinfo  {journal} {SoftwareX}\ }\textbf {\bibinfo {volume} {1}},\ \bibinfo {pages} {19} (\bibinfo {year} {2015})}\BibitemShut {NoStop}%
\bibitem [{\citenamefont {Phillips}\ \emph {et~al.}(2020)\citenamefont {Phillips}, \citenamefont {Hardy}, \citenamefont {Maia}, \citenamefont {Stone}, \citenamefont {Ribeiro}, \citenamefont {Bernardi}, \citenamefont {Buch}, \citenamefont {Fiorin}, \citenamefont {Hénin}, \citenamefont {Jiang}, \citenamefont {McGreevy}, \citenamefont {Melo}, \citenamefont {Radak}, \citenamefont {Skeel}, \citenamefont {Singharoy}, \citenamefont {Wang}, \citenamefont {Roux}, \citenamefont {Aksimentiev}, \citenamefont {Luthey-Schulten}, \citenamefont {Kalé}, \citenamefont {Schulten}, \citenamefont {Chipot},\ and\ \citenamefont {Tajkhorshid}}]{NAMD:2020}%
  \BibitemOpen
  \bibfield  {author} {\bibinfo {author} {\bibfnamefont {J.~C.}\ \bibnamefont {Phillips}}, \bibinfo {author} {\bibfnamefont {D.~J.}\ \bibnamefont {Hardy}}, \bibinfo {author} {\bibfnamefont {J.~D.~C.}\ \bibnamefont {Maia}}, \bibinfo {author} {\bibfnamefont {J.~E.}\ \bibnamefont {Stone}}, \bibinfo {author} {\bibfnamefont {J.~V.}\ \bibnamefont {Ribeiro}}, \bibinfo {author} {\bibfnamefont {R.~C.}\ \bibnamefont {Bernardi}}, \bibinfo {author} {\bibfnamefont {R.}~\bibnamefont {Buch}}, \bibinfo {author} {\bibfnamefont {G.}~\bibnamefont {Fiorin}}, \bibinfo {author} {\bibfnamefont {J.}~\bibnamefont {Hénin}}, \bibinfo {author} {\bibfnamefont {W.}~\bibnamefont {Jiang}}, \bibinfo {author} {\bibfnamefont {R.}~\bibnamefont {McGreevy}}, \bibinfo {author} {\bibfnamefont {M.~C.~R.}\ \bibnamefont {Melo}}, \bibinfo {author} {\bibfnamefont {B.~K.}\ \bibnamefont {Radak}}, \bibinfo {author} {\bibfnamefont {R.~D.}\ \bibnamefont {Skeel}}, \bibinfo {author} {\bibfnamefont {A.}~\bibnamefont {Singharoy}}, \bibinfo {author}
  {\bibfnamefont {Y.}~\bibnamefont {Wang}}, \bibinfo {author} {\bibfnamefont {B.}~\bibnamefont {Roux}}, \bibinfo {author} {\bibfnamefont {A.}~\bibnamefont {Aksimentiev}}, \bibinfo {author} {\bibfnamefont {Z.}~\bibnamefont {Luthey-Schulten}}, \bibinfo {author} {\bibfnamefont {L.~V.}\ \bibnamefont {Kalé}}, \bibinfo {author} {\bibfnamefont {K.}~\bibnamefont {Schulten}}, \bibinfo {author} {\bibfnamefont {C.}~\bibnamefont {Chipot}}, \ and\ \bibinfo {author} {\bibfnamefont {E.}~\bibnamefont {Tajkhorshid}},\ }\href@noop {} {\bibfield  {journal} {\bibinfo  {journal} {The Journal of Chemical Physics}\ }\textbf {\bibinfo {volume} {153}},\ \bibinfo {pages} {044130} (\bibinfo {year} {2020})}\BibitemShut {NoStop}%
\bibitem [{\citenamefont {Thompson}\ \emph {et~al.}(2022)\citenamefont {Thompson}, \citenamefont {Aktulga}, \citenamefont {Berger}, \citenamefont {Bolintineanu}, \citenamefont {Brown}, \citenamefont {Crozier}, \citenamefont {In't~Veld}, \citenamefont {Kohlmeyer}, \citenamefont {Moore}, \citenamefont {Nguyen} \emph {et~al.}}]{thompson2022lammps}%
  \BibitemOpen
  \bibfield  {author} {\bibinfo {author} {\bibfnamefont {A.~P.}\ \bibnamefont {Thompson}}, \bibinfo {author} {\bibfnamefont {H.~M.}\ \bibnamefont {Aktulga}}, \bibinfo {author} {\bibfnamefont {R.}~\bibnamefont {Berger}}, \bibinfo {author} {\bibfnamefont {D.~S.}\ \bibnamefont {Bolintineanu}}, \bibinfo {author} {\bibfnamefont {W.~M.}\ \bibnamefont {Brown}}, \bibinfo {author} {\bibfnamefont {P.~S.}\ \bibnamefont {Crozier}}, \bibinfo {author} {\bibfnamefont {P.~J.}\ \bibnamefont {In't~Veld}}, \bibinfo {author} {\bibfnamefont {A.}~\bibnamefont {Kohlmeyer}}, \bibinfo {author} {\bibfnamefont {S.~G.}\ \bibnamefont {Moore}}, \bibinfo {author} {\bibfnamefont {T.~D.}\ \bibnamefont {Nguyen}},  \emph {et~al.},\ }\href@noop {} {\bibfield  {journal} {\bibinfo  {journal} {Computer physics communications}\ }\textbf {\bibinfo {volume} {271}},\ \bibinfo {pages} {108171} (\bibinfo {year} {2022})}\BibitemShut {NoStop}%
\bibitem [{\citenamefont {Eastman}\ \emph {et~al.}(2017)\citenamefont {Eastman}, \citenamefont {Swails}, \citenamefont {Chodera}, \citenamefont {McGibbon}, \citenamefont {Zhao}, \citenamefont {Beauchamp}, \citenamefont {Wang}, \citenamefont {Simmonett}, \citenamefont {Harrigan}, \citenamefont {Stern} \emph {et~al.}}]{eastman2017openmm}%
  \BibitemOpen
  \bibfield  {author} {\bibinfo {author} {\bibfnamefont {P.}~\bibnamefont {Eastman}}, \bibinfo {author} {\bibfnamefont {J.}~\bibnamefont {Swails}}, \bibinfo {author} {\bibfnamefont {J.~D.}\ \bibnamefont {Chodera}}, \bibinfo {author} {\bibfnamefont {R.~T.}\ \bibnamefont {McGibbon}}, \bibinfo {author} {\bibfnamefont {Y.}~\bibnamefont {Zhao}}, \bibinfo {author} {\bibfnamefont {K.~A.}\ \bibnamefont {Beauchamp}}, \bibinfo {author} {\bibfnamefont {L.-P.}\ \bibnamefont {Wang}}, \bibinfo {author} {\bibfnamefont {A.~C.}\ \bibnamefont {Simmonett}}, \bibinfo {author} {\bibfnamefont {M.~P.}\ \bibnamefont {Harrigan}}, \bibinfo {author} {\bibfnamefont {C.~D.}\ \bibnamefont {Stern}},  \emph {et~al.},\ }\href@noop {} {\bibfield  {journal} {\bibinfo  {journal} {PLoS computational biology}\ }\textbf {\bibinfo {volume} {13}},\ \bibinfo {pages} {e1005659} (\bibinfo {year} {2017})}\BibitemShut {NoStop}%
\bibitem [{\citenamefont {Ayala}\ \emph {et~al.}(2022)\citenamefont {Ayala}, \citenamefont {Tomov}, \citenamefont {Stoyanov}, \citenamefont {Haidar},\ and\ \citenamefont {Dongarra}}]{Ayala:2022}%
  \BibitemOpen
  \bibfield  {author} {\bibinfo {author} {\bibfnamefont {A.}~\bibnamefont {Ayala}}, \bibinfo {author} {\bibfnamefont {S.}~\bibnamefont {Tomov}}, \bibinfo {author} {\bibfnamefont {M.}~\bibnamefont {Stoyanov}}, \bibinfo {author} {\bibfnamefont {A.}~\bibnamefont {Haidar}}, \ and\ \bibinfo {author} {\bibfnamefont {J.}~\bibnamefont {Dongarra}},\ }in\ \href@noop {} {\emph {\bibinfo {booktitle} {2022 IEEE International Parallel and Distributed Processing Symposium Workshops (IPDPSW)}}}\ (\bibinfo {year} {2022})\ pp.\ \bibinfo {pages} {372--381}\BibitemShut {NoStop}%
\bibitem [{\citenamefont {Zhou}\ \emph {et~al.}(2022)\citenamefont {Zhou}, \citenamefont {Kousha}, \citenamefont {Anthony}, \citenamefont {Shafie~Khorassani}, \citenamefont {Shafi}, \citenamefont {Subramoni},\ and\ \citenamefont {Panda}}]{Zhou:2022}%
  \BibitemOpen
  \bibfield  {author} {\bibinfo {author} {\bibfnamefont {Q.}~\bibnamefont {Zhou}}, \bibinfo {author} {\bibfnamefont {P.}~\bibnamefont {Kousha}}, \bibinfo {author} {\bibfnamefont {Q.}~\bibnamefont {Anthony}}, \bibinfo {author} {\bibfnamefont {K.}~\bibnamefont {Shafie~Khorassani}}, \bibinfo {author} {\bibfnamefont {A.}~\bibnamefont {Shafi}}, \bibinfo {author} {\bibfnamefont {H.}~\bibnamefont {Subramoni}}, \ and\ \bibinfo {author} {\bibfnamefont {D.~K.}\ \bibnamefont {Panda}},\ }in\ \href@noop {} {\emph {\bibinfo {booktitle} {High Performance Computing}}},\ \bibinfo {editor} {edited by\ \bibinfo {editor} {\bibfnamefont {A.-L.}\ \bibnamefont {Varbanescu}}, \bibinfo {editor} {\bibfnamefont {A.}~\bibnamefont {Bhatele}}, \bibinfo {editor} {\bibfnamefont {P.}~\bibnamefont {Luszczek}}, \ and\ \bibinfo {editor} {\bibfnamefont {B.}~\bibnamefont {Marc}}}\ (\bibinfo  {publisher} {Springer International Publishing},\ \bibinfo {address} {Cham},\ \bibinfo {year} {2022})\ pp.\ \bibinfo {pages} {3--25}\BibitemShut {NoStop}%
\bibitem [{\citenamefont {George}\ \emph {et~al.}(2022)\citenamefont {George}, \citenamefont {Mondal}, \citenamefont {Purnaprajna},\ and\ \citenamefont {Athri}}]{George:2022}%
  \BibitemOpen
  \bibfield  {author} {\bibinfo {author} {\bibfnamefont {A.}~\bibnamefont {George}}, \bibinfo {author} {\bibfnamefont {S.}~\bibnamefont {Mondal}}, \bibinfo {author} {\bibfnamefont {M.}~\bibnamefont {Purnaprajna}}, \ and\ \bibinfo {author} {\bibfnamefont {P.}~\bibnamefont {Athri}},\ }\href {\doibase 10.1021/acsomega.2c03189} {\bibfield  {journal} {\bibinfo  {journal} {ACS Omega}\ }\textbf {\bibinfo {volume} {7}},\ \bibinfo {pages} {32877} (\bibinfo {year} {2022})}\BibitemShut {NoStop}%
\bibitem [{\citenamefont {Gumerov}\ and\ \citenamefont {Duraiswami}(2004)}]{Gumerov:2004}%
  \BibitemOpen
  \bibfield  {author} {\bibinfo {author} {\bibfnamefont {N.~A.}\ \bibnamefont {Gumerov}}\ and\ \bibinfo {author} {\bibfnamefont {R.}~\bibnamefont {Duraiswami}},\ }\href@noop {} {\emph {\bibinfo {title} {Fast Multipole Methods for the Helmholtz Equation in Three Dimensions}}}\ (\bibinfo  {publisher} {Elsevier},\ \bibinfo {year} {2004})\BibitemShut {NoStop}%
\bibitem [{\citenamefont {Fennell}\ and\ \citenamefont {Gezelter}(2006)}]{Fennell:2006}%
  \BibitemOpen
  \bibfield  {author} {\bibinfo {author} {\bibfnamefont {C.~J.}\ \bibnamefont {Fennell}}\ and\ \bibinfo {author} {\bibfnamefont {J.~D.}\ \bibnamefont {Gezelter}},\ }\href@noop {} {\bibfield  {journal} {\bibinfo  {journal} {The Journal of Chemical Physics}\ }\textbf {\bibinfo {volume} {124}},\ \bibinfo {pages} {234104} (\bibinfo {year} {2006})}\BibitemShut {NoStop}%
\bibitem [{\citenamefont {Rodgers}\ and\ \citenamefont {Weeks}(2008)}]{Rodgers:2008}%
  \BibitemOpen
  \bibfield  {author} {\bibinfo {author} {\bibfnamefont {J.~M.}\ \bibnamefont {Rodgers}}\ and\ \bibinfo {author} {\bibfnamefont {J.~D.}\ \bibnamefont {Weeks}},\ }\href@noop {} {\bibfield  {journal} {\bibinfo  {journal} {Proceedings of the National Academy of Sciences}\ }\textbf {\bibinfo {volume} {105}},\ \bibinfo {pages} {19136} (\bibinfo {year} {2008})}\BibitemShut {NoStop}%
\bibitem [{\citenamefont {Dehnen}(2002)}]{Dehnen:2002}%
  \BibitemOpen
  \bibfield  {author} {\bibinfo {author} {\bibfnamefont {W.}~\bibnamefont {Dehnen}},\ }\href@noop {} {\bibfield  {journal} {\bibinfo  {journal} {Journal of Computational Physics}\ }\textbf {\bibinfo {volume} {179}},\ \bibinfo {pages} {27} (\bibinfo {year} {2002})}\BibitemShut {NoStop}%
\bibitem [{\citenamefont {Boateng}(2019)}]{Boateng-2019}%
  \BibitemOpen
  \bibfield  {author} {\bibinfo {author} {\bibfnamefont {H.~A.}\ \bibnamefont {Boateng}},\ }\href@noop {} {\bibfield  {journal} {\bibinfo  {journal} {Journal of Chemical Theory and Computation}\ }\textbf {\bibinfo {volume} {16}},\ \bibinfo {pages} {7} (\bibinfo {year} {2019})}\BibitemShut {NoStop}%
\bibitem [{\citenamefont {Skeel}, \citenamefont {Tezcan},\ and\ \citenamefont {Hardy}(2002)}]{Skeel:2002}%
  \BibitemOpen
  \bibfield  {author} {\bibinfo {author} {\bibfnamefont {R.~D.}\ \bibnamefont {Skeel}}, \bibinfo {author} {\bibfnamefont {I.}~\bibnamefont {Tezcan}}, \ and\ \bibinfo {author} {\bibfnamefont {D.~J.}\ \bibnamefont {Hardy}},\ }\href@noop {} {\bibfield  {journal} {\bibinfo  {journal} {Journal of Computational Chemistry}\ }\textbf {\bibinfo {volume} {23}},\ \bibinfo {pages} {673} (\bibinfo {year} {2002})}\BibitemShut {NoStop}%
\bibitem [{\citenamefont {Hardy}\ \emph {et~al.}(2015)\citenamefont {Hardy}, \citenamefont {Wu}, \citenamefont {Phillips}, \citenamefont {Stone}, \citenamefont {Skeel},\ and\ \citenamefont {Schulten}}]{Hardy:2014}%
  \BibitemOpen
  \bibfield  {author} {\bibinfo {author} {\bibfnamefont {D.~J.}\ \bibnamefont {Hardy}}, \bibinfo {author} {\bibfnamefont {Z.}~\bibnamefont {Wu}}, \bibinfo {author} {\bibfnamefont {J.~C.}\ \bibnamefont {Phillips}}, \bibinfo {author} {\bibfnamefont {J.~E.}\ \bibnamefont {Stone}}, \bibinfo {author} {\bibfnamefont {R.~D.}\ \bibnamefont {Skeel}}, \ and\ \bibinfo {author} {\bibfnamefont {K.}~\bibnamefont {Schulten}},\ }\href@noop {} {\bibfield  {journal} {\bibinfo  {journal} {Journal of Chemical Theory and Computation}\ }\textbf {\bibinfo {volume} {11}},\ \bibinfo {pages} {766} (\bibinfo {year} {2015})}\BibitemShut {NoStop}%
\bibitem [{\citenamefont {{Hockney}}\ and\ \citenamefont {{Eastwood}}(1981)}]{Hockney:1981}%
  \BibitemOpen
  \bibfield  {author} {\bibinfo {author} {\bibfnamefont {R.~W.}\ \bibnamefont {{Hockney}}}\ and\ \bibinfo {author} {\bibfnamefont {J.~W.}\ \bibnamefont {{Eastwood}}},\ }\href@noop {} {\emph {\bibinfo {title} {{Computer Simulation Using Particles}}}}\ (\bibinfo  {publisher} {McGraw-Hill},\ \bibinfo {year} {1981})\BibitemShut {NoStop}%
\bibitem [{\citenamefont {Beckers}, \citenamefont {Lowe},\ and\ \citenamefont {Leeuw}(1998)}]{Beckers:1998}%
  \BibitemOpen
  \bibfield  {author} {\bibinfo {author} {\bibfnamefont {J.~V.~L.}\ \bibnamefont {Beckers}}, \bibinfo {author} {\bibfnamefont {C.~P.}\ \bibnamefont {Lowe}}, \ and\ \bibinfo {author} {\bibfnamefont {S.~W.~D.}\ \bibnamefont {Leeuw}},\ }\href {\doibase 10.1080/08927029808022044} {\bibfield  {journal} {\bibinfo  {journal} {Molecular Simulation}\ }\textbf {\bibinfo {volume} {20}},\ \bibinfo {pages} {369} (\bibinfo {year} {1998})}\BibitemShut {NoStop}%
\bibitem [{\citenamefont {Sagui}\ and\ \citenamefont {Darden}(2001)}]{Sagui:2001}%
  \BibitemOpen
  \bibfield  {author} {\bibinfo {author} {\bibfnamefont {C.}~\bibnamefont {Sagui}}\ and\ \bibinfo {author} {\bibfnamefont {T.}~\bibnamefont {Darden}},\ }\href@noop {} {\bibfield  {journal} {\bibinfo  {journal} {The Journal of Chemical Physics}\ }\textbf {\bibinfo {volume} {114}},\ \bibinfo {pages} {6578} (\bibinfo {year} {2001})}\BibitemShut {NoStop}%
\bibitem [{\citenamefont {Rottler}\ and\ \citenamefont {Maggs}(2004{\natexlab{a}})}]{Rottler:2004}%
  \BibitemOpen
  \bibfield  {author} {\bibinfo {author} {\bibfnamefont {J.}~\bibnamefont {Rottler}}\ and\ \bibinfo {author} {\bibfnamefont {A.~C.}\ \bibnamefont {Maggs}},\ }\href@noop {} {\bibfield  {journal} {\bibinfo  {journal} {Phys. Rev. Lett.}\ }\textbf {\bibinfo {volume} {93}},\ \bibinfo {pages} {170201} (\bibinfo {year} {2004}{\natexlab{a}})}\BibitemShut {NoStop}%
\bibitem [{\citenamefont {Ryckaert}, \citenamefont {Bellemans},\ and\ \citenamefont {and}(1981)}]{Ryckaert:1981}%
  \BibitemOpen
  \bibfield  {author} {\bibinfo {author} {\bibfnamefont {J.-P.}\ \bibnamefont {Ryckaert}}, \bibinfo {author} {\bibfnamefont {A.}~\bibnamefont {Bellemans}}, \ and\ \bibinfo {author} {\bibfnamefont {G.~C.}\ \bibnamefont {and}},\ }\href {\doibase 10.1080/00268978100102931} {\bibfield  {journal} {\bibinfo  {journal} {Molecular Physics}\ }\textbf {\bibinfo {volume} {44}},\ \bibinfo {pages} {979} (\bibinfo {year} {1981})}\BibitemShut {NoStop}%
\bibitem [{\citenamefont {Ryckaert}, \citenamefont {Ciccotti},\ and\ \citenamefont {Berendsen}(1977)}]{ryckaert:1977}%
  \BibitemOpen
  \bibfield  {author} {\bibinfo {author} {\bibfnamefont {J.-P.}\ \bibnamefont {Ryckaert}}, \bibinfo {author} {\bibfnamefont {G.}~\bibnamefont {Ciccotti}}, \ and\ \bibinfo {author} {\bibfnamefont {H.~J.}\ \bibnamefont {Berendsen}},\ }\href@noop {} {\bibfield  {journal} {\bibinfo  {journal} {Journal of Computational Physics}\ }\textbf {\bibinfo {volume} {23}},\ \bibinfo {pages} {327} (\bibinfo {year} {1977})}\BibitemShut {NoStop}%
\bibitem [{\citenamefont {Coretti}, \citenamefont {Bonella},\ and\ \citenamefont {Ciccotti}(2018)}]{coretti:2018b}%
  \BibitemOpen
  \bibfield  {author} {\bibinfo {author} {\bibfnamefont {A.}~\bibnamefont {Coretti}}, \bibinfo {author} {\bibfnamefont {S.}~\bibnamefont {Bonella}}, \ and\ \bibinfo {author} {\bibfnamefont {G.}~\bibnamefont {Ciccotti}},\ }\href@noop {} {\bibfield  {journal} {\bibinfo  {journal} {The Journal of Chemical Physics}\ }\textbf {\bibinfo {volume} {149}},\ \bibinfo {pages} {191102} (\bibinfo {year} {2018})}\BibitemShut {NoStop}%
\bibitem [{\citenamefont {Girardier}\ \emph {et~al.}(2021)\citenamefont {Girardier}, \citenamefont {Coretti}, \citenamefont {Ciccotti},\ and\ \citenamefont {Bonella}}]{girardier:2021}%
  \BibitemOpen
  \bibfield  {author} {\bibinfo {author} {\bibfnamefont {D.}~\bibnamefont {Girardier}}, \bibinfo {author} {\bibfnamefont {A.}~\bibnamefont {Coretti}}, \bibinfo {author} {\bibfnamefont {G.}~\bibnamefont {Ciccotti}}, \ and\ \bibinfo {author} {\bibfnamefont {S.}~\bibnamefont {Bonella}},\ }\href@noop {} {\bibfield  {journal} {\bibinfo  {journal} {The European Physical Journal B}\ }\textbf {\bibinfo {volume} {94}},\ \bibinfo {pages} {158} (\bibinfo {year} {2021})}\BibitemShut {NoStop}%
\bibitem [{\citenamefont {Coretti}\ \emph {et~al.}(2020)\citenamefont {Coretti}, \citenamefont {Scalfi}, \citenamefont {Bacon}, \citenamefont {Rotenberg}, \citenamefont {Vuilleumier}, \citenamefont {Ciccotti}, \citenamefont {Salanne},\ and\ \citenamefont {Bonella}}]{coretti:2020b}%
  \BibitemOpen
  \bibfield  {author} {\bibinfo {author} {\bibfnamefont {A.}~\bibnamefont {Coretti}}, \bibinfo {author} {\bibfnamefont {L.}~\bibnamefont {Scalfi}}, \bibinfo {author} {\bibfnamefont {C.}~\bibnamefont {Bacon}}, \bibinfo {author} {\bibfnamefont {B.}~\bibnamefont {Rotenberg}}, \bibinfo {author} {\bibfnamefont {R.}~\bibnamefont {Vuilleumier}}, \bibinfo {author} {\bibfnamefont {G.}~\bibnamefont {Ciccotti}}, \bibinfo {author} {\bibfnamefont {M.}~\bibnamefont {Salanne}}, \ and\ \bibinfo {author} {\bibfnamefont {S.}~\bibnamefont {Bonella}},\ }\href@noop {} {\bibfield  {journal} {\bibinfo  {journal} {The Journal of Chemical Physics}\ }\textbf {\bibinfo {volume} {152}},\ \bibinfo {pages} {194701} (\bibinfo {year} {2020})}\BibitemShut {NoStop}%
\bibitem [{\citenamefont {Marin-Laflèche}\ \emph {et~al.}(2020)\citenamefont {Marin-Laflèche}, \citenamefont {Haefele}, \citenamefont {Scalfi}, \citenamefont {Coretti}, \citenamefont {Dufils}, \citenamefont {Jeanmairet}, \citenamefont {Reed}, \citenamefont {Serva}, \citenamefont {Berthin}, \citenamefont {Bacon}, \citenamefont {Bonella}, \citenamefont {Rotenberg}, \citenamefont {Madden},\ and\ \citenamefont {Salanne}}]{MW:2020}%
  \BibitemOpen
  \bibfield  {author} {\bibinfo {author} {\bibfnamefont {A.}~\bibnamefont {Marin-Laflèche}}, \bibinfo {author} {\bibfnamefont {M.}~\bibnamefont {Haefele}}, \bibinfo {author} {\bibfnamefont {L.}~\bibnamefont {Scalfi}}, \bibinfo {author} {\bibfnamefont {A.}~\bibnamefont {Coretti}}, \bibinfo {author} {\bibfnamefont {T.}~\bibnamefont {Dufils}}, \bibinfo {author} {\bibfnamefont {G.}~\bibnamefont {Jeanmairet}}, \bibinfo {author} {\bibfnamefont {S.~K.}\ \bibnamefont {Reed}}, \bibinfo {author} {\bibfnamefont {A.}~\bibnamefont {Serva}}, \bibinfo {author} {\bibfnamefont {R.}~\bibnamefont {Berthin}}, \bibinfo {author} {\bibfnamefont {C.}~\bibnamefont {Bacon}}, \bibinfo {author} {\bibfnamefont {S.}~\bibnamefont {Bonella}}, \bibinfo {author} {\bibfnamefont {B.}~\bibnamefont {Rotenberg}}, \bibinfo {author} {\bibfnamefont {P.~A.}\ \bibnamefont {Madden}}, \ and\ \bibinfo {author} {\bibfnamefont {M.}~\bibnamefont {Salanne}},\ }\href {\doibase 10.21105/joss.02373} {\bibfield  {journal} {\bibinfo  {journal} {Journal of Open
  Source Software}\ }\textbf {\bibinfo {volume} {5}},\ \bibinfo {pages} {2373} (\bibinfo {year} {2020})}\BibitemShut {NoStop}%
\bibitem [{\citenamefont {Bonella}\ \emph {et~al.}(2020)\citenamefont {Bonella}, \citenamefont {Coretti}, \citenamefont {Vuilleumier},\ and\ \citenamefont {Ciccotti}}]{bonella:2020}%
  \BibitemOpen
  \bibfield  {author} {\bibinfo {author} {\bibfnamefont {S.}~\bibnamefont {Bonella}}, \bibinfo {author} {\bibfnamefont {A.}~\bibnamefont {Coretti}}, \bibinfo {author} {\bibfnamefont {R.}~\bibnamefont {Vuilleumier}}, \ and\ \bibinfo {author} {\bibfnamefont {G.}~\bibnamefont {Ciccotti}},\ }\href@noop {} {\bibfield  {journal} {\bibinfo  {journal} {Phys. Chem. Chem. Phys.}\ }\textbf {\bibinfo {volume} {22}},\ \bibinfo {pages} {10775} (\bibinfo {year} {2020})}\BibitemShut {NoStop}%
\bibitem [{\citenamefont {Coretti}\ \emph {et~al.}(2022)\citenamefont {Coretti}, \citenamefont {Baird}, \citenamefont {Vuilleumier},\ and\ \citenamefont {Bonella}}]{coretti:2022}%
  \BibitemOpen
  \bibfield  {author} {\bibinfo {author} {\bibfnamefont {A.}~\bibnamefont {Coretti}}, \bibinfo {author} {\bibfnamefont {T.}~\bibnamefont {Baird}}, \bibinfo {author} {\bibfnamefont {R.}~\bibnamefont {Vuilleumier}}, \ and\ \bibinfo {author} {\bibfnamefont {S.}~\bibnamefont {Bonella}},\ }\href@noop {} {\bibfield  {journal} {\bibinfo  {journal} {The Journal of Chemical Physics}\ }\textbf {\bibinfo {volume} {157}} (\bibinfo {year} {2022})}\BibitemShut {NoStop}%
\bibitem [{\citenamefont {Im}, \citenamefont {Beglov},\ and\ \citenamefont {Roux}(1998)}]{Im:1998}%
  \BibitemOpen
  \bibfield  {author} {\bibinfo {author} {\bibfnamefont {W.}~\bibnamefont {Im}}, \bibinfo {author} {\bibfnamefont {D.}~\bibnamefont {Beglov}}, \ and\ \bibinfo {author} {\bibfnamefont {B.}~\bibnamefont {Roux}},\ }\href@noop {} {\bibfield  {journal} {\bibinfo  {journal} {Computer Physics Communications}\ }\textbf {\bibinfo {volume} {111}},\ \bibinfo {pages} {59} (\bibinfo {year} {1998})}\BibitemShut {NoStop}%
\bibitem [{Note1()}]{Note1}%
  \BibitemOpen
  \bibinfo {note} {The, standard, form of the matrix $\protect \bm {M}$, is obtained by first considering a discretization on a grid with indices $i, j, k = 0, \protect \ldots , N - 1$, where $N$ denotes the number of grid points along each Cartesian direction. The discrete representation of the Laplacian and the mapping $n = i + jN + kN^2$ are then used to express the matrix with only two indices $n,m = 0,...,\protect \mathcal {N}-1$, where $\protect \mathcal {N} = N^3$. $M_{nm}= -6\delta _{nm} + \delta _{n(m-1)} + \delta _{n(m+1)} + \delta _{n(m-N)} + \delta _{n(m+N)} + \delta _{n(m-N^2)} +\delta _{n(m+N^2)}$, with appropriate provisions to account for periodic boundary conditions. We recall, for future convenience, that this matrix is symmetrical and has only seven non-zero entries for row.}\BibitemShut {Stop}%
\bibitem [{Note2()}]{Note2}%
  \BibitemOpen
  \bibinfo {note} {In the context of \protect \textit {first principles} \ac {md}, this would be analogous to the Born-Oppenheimer approach - albeit with a different condition for the non ionic \ac {dof}.}\BibitemShut {Stop}%
\bibitem [{\citenamefont {Gholami}\ \emph {et~al.}(2016)\citenamefont {Gholami}, \citenamefont {Malhotra}, \citenamefont {Sundar},\ and\ \citenamefont {Biros}}]{gholami:2016}%
  \BibitemOpen
  \bibfield  {author} {\bibinfo {author} {\bibfnamefont {A.}~\bibnamefont {Gholami}}, \bibinfo {author} {\bibfnamefont {D.}~\bibnamefont {Malhotra}}, \bibinfo {author} {\bibfnamefont {H.}~\bibnamefont {Sundar}}, \ and\ \bibinfo {author} {\bibfnamefont {G.}~\bibnamefont {Biros}},\ }\href@noop {} {\bibfield  {journal} {\bibinfo  {journal} {SIAM Journal on Scientific Computing}\ }\textbf {\bibinfo {volume} {38}},\ \bibinfo {pages} {C280} (\bibinfo {year} {2016})}\BibitemShut {NoStop}%
\bibitem [{\citenamefont {Ibeid}, \citenamefont {Olson},\ and\ \citenamefont {Gropp}(2020)}]{ibeid:2020}%
  \BibitemOpen
  \bibfield  {author} {\bibinfo {author} {\bibfnamefont {H.}~\bibnamefont {Ibeid}}, \bibinfo {author} {\bibfnamefont {L.}~\bibnamefont {Olson}}, \ and\ \bibinfo {author} {\bibfnamefont {W.}~\bibnamefont {Gropp}},\ }\href@noop {} {\bibfield  {journal} {\bibinfo  {journal} {Journal of Parallel and Distributed Computing}\ }\textbf {\bibinfo {volume} {136}},\ \bibinfo {pages} {63} (\bibinfo {year} {2020})}\BibitemShut {NoStop}%
\bibitem [{\citenamefont {Gilson}\ \emph {et~al.}(1985)\citenamefont {Gilson}, \citenamefont {Rashin}, \citenamefont {Fine},\ and\ \citenamefont {Honig}}]{Gilson:1985}%
  \BibitemOpen
  \bibfield  {author} {\bibinfo {author} {\bibfnamefont {M.~K.}\ \bibnamefont {Gilson}}, \bibinfo {author} {\bibfnamefont {A.}~\bibnamefont {Rashin}}, \bibinfo {author} {\bibfnamefont {R.}~\bibnamefont {Fine}}, \ and\ \bibinfo {author} {\bibfnamefont {B.}~\bibnamefont {Honig}},\ }\href@noop {} {\bibfield  {journal} {\bibinfo  {journal} {Journal of Molecular Biology}\ }\textbf {\bibinfo {volume} {184}},\ \bibinfo {pages} {503} (\bibinfo {year} {1985})}\BibitemShut {NoStop}%
\bibitem [{\citenamefont {Gilson}\ \emph {et~al.}(1993)\citenamefont {Gilson}, \citenamefont {Davis}, \citenamefont {Luty},\ and\ \citenamefont {McCammon}}]{Gilson:1993}%
  \BibitemOpen
  \bibfield  {author} {\bibinfo {author} {\bibfnamefont {M.~K.}\ \bibnamefont {Gilson}}, \bibinfo {author} {\bibfnamefont {M.~E.}\ \bibnamefont {Davis}}, \bibinfo {author} {\bibfnamefont {B.~A.}\ \bibnamefont {Luty}}, \ and\ \bibinfo {author} {\bibfnamefont {J.~A.}\ \bibnamefont {McCammon}},\ }\href@noop {} {\bibfield  {journal} {\bibinfo  {journal} {The Journal of Physical Chemistry}\ }\textbf {\bibinfo {volume} {97}},\ \bibinfo {pages} {3591} (\bibinfo {year} {1993})}\BibitemShut {NoStop}%
\bibitem [{\citenamefont {Rottler}\ and\ \citenamefont {Maggs}(2004{\natexlab{b}})}]{Maggs:2004}%
  \BibitemOpen
  \bibfield  {author} {\bibinfo {author} {\bibfnamefont {J.}~\bibnamefont {Rottler}}\ and\ \bibinfo {author} {\bibfnamefont {A.~C.}\ \bibnamefont {Maggs}},\ }\href@noop {} {\bibfield  {journal} {\bibinfo  {journal} {The Journal of Chemical Physics}\ }\textbf {\bibinfo {volume} {120}},\ \bibinfo {pages} {3119} (\bibinfo {year} {2004}{\natexlab{b}})}\BibitemShut {NoStop}%
\bibitem [{\citenamefont {Galamba}\ and\ \citenamefont {Costa~Cabral}(2007)}]{Galamba:2007}%
  \BibitemOpen
  \bibfield  {author} {\bibinfo {author} {\bibfnamefont {N.}~\bibnamefont {Galamba}}\ and\ \bibinfo {author} {\bibfnamefont {B.~J.}\ \bibnamefont {Costa~Cabral}},\ }\href@noop {} {\bibfield  {journal} {\bibinfo  {journal} {The Journal of Chemical Physics}\ }\textbf {\bibinfo {volume} {126}},\ \bibinfo {pages} {124502} (\bibinfo {year} {2007})}\BibitemShut {NoStop}%
\bibitem [{\citenamefont {F.~Mouhat}\ and\ \citenamefont {Pierleoni}(2013)}]{Mouhat:2013}%
  \BibitemOpen
  \bibfield  {author} {\bibinfo {author} {\bibfnamefont {S.~B.}\ \bibnamefont {F.~Mouhat}}\ and\ \bibinfo {author} {\bibfnamefont {C.}~\bibnamefont {Pierleoni}},\ }\href {https://doi.org/10.1080/00268976.2013.846486} {\bibfield  {journal} {\bibinfo  {journal} {Molecular Physics}\ }\textbf {\bibinfo {volume} {111}},\ \bibinfo {pages} {3651} (\bibinfo {year} {2013})}\BibitemShut {NoStop}%
\bibitem [{\citenamefont {Tosi}\ and\ \citenamefont {Fumi}(1964)}]{Tosi:1964}%
  \BibitemOpen
  \bibfield  {author} {\bibinfo {author} {\bibfnamefont {M.}~\bibnamefont {Tosi}}\ and\ \bibinfo {author} {\bibfnamefont {F.}~\bibnamefont {Fumi}},\ }\href@noop {} {\bibfield  {journal} {\bibinfo  {journal} {Journal of Physics and Chemistry of Solids}\ }\textbf {\bibinfo {volume} {25}},\ \bibinfo {pages} {45} (\bibinfo {year} {1964})}\BibitemShut {NoStop}%
\bibitem [{\citenamefont {Sivak}, \citenamefont {Chodera},\ and\ \citenamefont {Crooks}(2013)}]{Sivak:2013}%
  \BibitemOpen
  \bibfield  {author} {\bibinfo {author} {\bibfnamefont {D.~A.}\ \bibnamefont {Sivak}}, \bibinfo {author} {\bibfnamefont {J.~D.}\ \bibnamefont {Chodera}}, \ and\ \bibinfo {author} {\bibfnamefont {G.~E.}\ \bibnamefont {Crooks}},\ }\href@noop {} {\bibfield  {journal} {\bibinfo  {journal} {The Journal of Physical Chemistry. B}\ }\textbf {\bibinfo {volume} {118}},\ \bibinfo {pages} {6466 } (\bibinfo {year} {2013})}\BibitemShut {NoStop}%
\bibitem [{Note3()}]{Note3}%
  \BibitemOpen
  \bibinfo {note} {This implementation is quite certainly not as efficient as \ac {mg} functions in high end community codes, but since here we are interested in relative performance of \ac {pmaze} and \ac {mg} at equal implementation level, it is sufficient for our purposes. We also note that the \ac {mg} function called is the same, with different input, for the direct solution of the Poisson equation and for \ac {pmaze}. This implies that transfer of our approach to community codes requires only the set up of the extended dynamical system, and can directly benefit from existing optimizations.}\BibitemShut {Stop}%
\bibitem [{\citenamefont {Briggs}, \citenamefont {Henson},\ and\ \citenamefont {McCormick}(2000)}]{briggs:2000}%
  \BibitemOpen
  \bibfield  {author} {\bibinfo {author} {\bibfnamefont {W.~L.}\ \bibnamefont {Briggs}}, \bibinfo {author} {\bibfnamefont {V.~E.}\ \bibnamefont {Henson}}, \ and\ \bibinfo {author} {\bibfnamefont {S.~F.}\ \bibnamefont {McCormick}},\ }\href@noop {} {\emph {\bibinfo {title} {A Multigrid Tutorial}}},\ \bibinfo {edition} {2nd}\ ed.\ (\bibinfo  {publisher} {SIAM},\ \bibinfo {year} {2000})\BibitemShut {NoStop}%
\bibitem [{\citenamefont {Leimkuhler}(1999)}]{leimkuhler1999comparison}%
  \BibitemOpen
  \bibfield  {author} {\bibinfo {author} {\bibfnamefont {B.~J.}\ \bibnamefont {Leimkuhler}},\ }in\ \href@noop {} {\emph {\bibinfo {booktitle} {Computational Molecular Dynamics: Challenges, Methods, Ideas: Proceedings of the 2nd International Symposium on Algorithms for Macromolecular Modelling, Berlin, May 21--24, 1997}}}\ (\bibinfo {organization} {Springer},\ \bibinfo {year} {1999})\ pp.\ \bibinfo {pages} {349--362}\BibitemShut {NoStop}%
\bibitem [{\citenamefont {Klapper}\ \emph {et~al.}(1986)\citenamefont {Klapper}, \citenamefont {Hagstrom}, \citenamefont {Fine}, \citenamefont {Sharp},\ and\ \citenamefont {Honig}}]{Klapper:1986}%
  \BibitemOpen
  \bibfield  {author} {\bibinfo {author} {\bibfnamefont {I.}~\bibnamefont {Klapper}}, \bibinfo {author} {\bibfnamefont {R.}~\bibnamefont {Hagstrom}}, \bibinfo {author} {\bibfnamefont {R.}~\bibnamefont {Fine}}, \bibinfo {author} {\bibfnamefont {K.}~\bibnamefont {Sharp}}, \ and\ \bibinfo {author} {\bibfnamefont {B.}~\bibnamefont {Honig}},\ }\href@noop {} {\bibfield  {journal} {\bibinfo  {journal} {Proteins: Structure, Function, and Bioinformatics}\ }\textbf {\bibinfo {volume} {1}},\ \bibinfo {pages} {47} (\bibinfo {year} {1986})}\BibitemShut {NoStop}%
\bibitem [{\citenamefont {Nicholls}\ and\ \citenamefont {Honig}(1991)}]{Nicholls:1991}%
  \BibitemOpen
  \bibfield  {author} {\bibinfo {author} {\bibfnamefont {A.}~\bibnamefont {Nicholls}}\ and\ \bibinfo {author} {\bibfnamefont {B.}~\bibnamefont {Honig}},\ }\href@noop {} {\bibfield  {journal} {\bibinfo  {journal} {Journal of Computational Chemistry}\ }\textbf {\bibinfo {volume} {12}},\ \bibinfo {pages} {435} (\bibinfo {year} {1991})}\BibitemShut {NoStop}%
\bibitem [{\citenamefont {Jo}\ \emph {et~al.}(2008)\citenamefont {Jo}, \citenamefont {Kim}, \citenamefont {Iyer},\ and\ \citenamefont {Im}}]{CHARMM-GUI:2008}%
  \BibitemOpen
  \bibfield  {author} {\bibinfo {author} {\bibfnamefont {S.}~\bibnamefont {Jo}}, \bibinfo {author} {\bibfnamefont {T.}~\bibnamefont {Kim}}, \bibinfo {author} {\bibfnamefont {V.~G.}\ \bibnamefont {Iyer}}, \ and\ \bibinfo {author} {\bibfnamefont {W.}~\bibnamefont {Im}},\ }\href@noop {} {\bibfield  {journal} {\bibinfo  {journal} {Journal of Computational Chemistry}\ }\textbf {\bibinfo {volume} {29}},\ \bibinfo {pages} {1859} (\bibinfo {year} {2008})}\BibitemShut {NoStop}%
\bibitem [{\citenamefont {Hestenes}\ and\ \citenamefont {Stiefel}(1952)}]{Hestenes:1952}%
  \BibitemOpen
  \bibfield  {author} {\bibinfo {author} {\bibfnamefont {M.~R.}\ \bibnamefont {Hestenes}}\ and\ \bibinfo {author} {\bibfnamefont {E.}~\bibnamefont {Stiefel}},\ }\href@noop {} {\bibfield  {journal} {\bibinfo  {journal} {Journal of research of the National Bureau of Standards}\ }\textbf {\bibinfo {volume} {49}},\ \bibinfo {pages} {409} (\bibinfo {year} {1952})}\BibitemShut {NoStop}%
\bibitem [{\citenamefont {Elber}, \citenamefont {Ruymgaart},\ and\ \citenamefont {Hess}(2011)}]{elber:2011}%
  \BibitemOpen
  \bibfield  {author} {\bibinfo {author} {\bibfnamefont {R.}~\bibnamefont {Elber}}, \bibinfo {author} {\bibfnamefont {A.~P.}\ \bibnamefont {Ruymgaart}}, \ and\ \bibinfo {author} {\bibfnamefont {B.}~\bibnamefont {Hess}},\ }\href@noop {} {\bibfield  {journal} {\bibinfo  {journal} {The European Physical Journal Special Topics}\ }\textbf {\bibinfo {volume} {200}},\ \bibinfo {pages} {211} (\bibinfo {year} {2011})}\BibitemShut {NoStop}%
\bibitem [{\citenamefont {Polyak}(1987)}]{polyak:1987}%
  \BibitemOpen
  \bibfield  {author} {\bibinfo {author} {\bibfnamefont {B.~T.}\ \bibnamefont {Polyak}},\ }\href@noop {} {\emph {\bibinfo {title} {Introduction to optimization}}}\ (\bibinfo  {publisher} {New York, Optimization Software},\ \bibinfo {year} {1987})\BibitemShut {NoStop}%
\bibitem [{\citenamefont {Beckermann}\ and\ \citenamefont {Kuijlaars}(2002)}]{Beckermann:2002}%
  \BibitemOpen
  \bibfield  {author} {\bibinfo {author} {\bibfnamefont {B.}~\bibnamefont {Beckermann}}\ and\ \bibinfo {author} {\bibfnamefont {A.~B.}\ \bibnamefont {Kuijlaars}},\ }\href@noop {} {\bibfield  {journal} {\bibinfo  {journal} {ETNA. Electronic Transactions on Numerical Analysis [electronic only]}\ }\textbf {\bibinfo {volume} {14}},\ \bibinfo {pages} {1} (\bibinfo {year} {2002})}\BibitemShut {NoStop}%
\end{thebibliography}%

\end{document}